\documentclass[fleqn,usenatbib]{mnras}

\usepackage{newtxtext,newtxmath}
\usepackage{pdflscape}

\usepackage[T1]{fontenc}

\DeclareRobustCommand{\VAN}[3]{#2}
\let\VANthebibliography\thebibliography
\def\thebibliography{\DeclareRobustCommand{\VAN}[3]{##3}\VANthebibliography}

\usepackage{graphicx}	
\usepackage{amsmath}	
\usepackage{orcidlink}
\usepackage{anyfontsize}

\newcommand{\tiu}{\,J\,m$^{-2}$\,K$^{-1}$\,s$^{-1/2}$}	

\definecolor{revone}{RGB}{0,0,0}
\definecolor{revtwo}{RGB}{0,0,0}
\definecolor{editor}{RGB}{0,0,0}
\definecolor{author}{RGB}{0,0,0}

\title[TEMPEST: A Modular Rough Surface Python TPM]{TEMPEST - A modular Python-based thermophysical model for rough, irregularly-shaped airless bodies}

\author[D. G. Lyster and C. J. A. Howett]{
Duncan G. Lyster\,\orcidlink{0009-0006-3295-7685},$^{1}$\thanks{E-mail: duncan.lyster@physics.ox.ac.uk (DGL)}
Carly J. A. Howett\,\orcidlink{0000-0003-1869-4947},$^{1}$
\\
$^{1}$Atmospheric, Oceanic \& Planetary Physics, Clarendon Laboratory, Sherrington Rd, Oxford OX1 3PU, UK\\
}

\date{\textit{This is a pre-copyedited, author-produced PDF of an article accepted for publication in RAS Techniques and Instruments following peer review.}}

\pubyear{\the\year{}}

\begin{document}
\label{firstpage}
\pagerange{\pageref{firstpage}--\pageref{lastpage}}
\maketitle

\begin{abstract}
Comparing observed and predicted surface temperatures on airless planetary bodies is crucial for interpreting their surface processes, constraining activity, and deriving their surface properties. We present a new tool in this endeavour: TEMPEST (Thermophysical Evolution Model for Planetary Environment Surface Temperatures). It is a modular, open-source Python model for diurnal thermal evolution on irregular bodies. TEMPEST can reproduce non-Lambertian multiple scattering using arbitrary Bidirectional Reflectance Distribution Functions (BRDFs) and notably incorporates surface roughness \textcolor{revone}{on sub-facet scales} through representative sphere-section geometry. The model includes an explicit time-stepping one-dimensional periodic conduction solver influenced by Spencer's thermal model Thermprojrs, and a semi-implicit Crank-Nicolson solver, and can be paired with other thermal solvers. Precomputed look-up tables enable fast application of scattering laws (Lambertian, Lommel-Seeliger, Hapke, or input empirical BRDF data), allowing the user to assess their effects in permanently shadowed regions or partially shadowed terrains. TEMPEST is parallelised, running on personal computers in minutes for thousands of facets and scaling to high-performance servers for models exceeding 100,000 facets. Validation was performed against Thermprojrs in one dimension, against Sorli's Binary Thermophysical Model in three dimensions, and with empirical thermal data from the Lunar Reconnaissance Orbiter, showing good agreement in each case. Application to the high-resolution shape-model of 25143 Itokawa shows the importance of self-shadowing and radiative mutual heating in controlling local temperatures. TEMPEST has already been used to analyse data from the Lucy Mission, Galileo, and ground-based observations, and is adaptable to upcoming datasets from targets of missions such as Comet Interceptor and Europa Clipper.
\end{abstract}




Space mission infrared (IR) data, such as those produced by multispectral thermal mapping instruments, are valuable on their own, but much more information about the observed surfaces can be extracted through comparison of their results with those of a thermophysical model (TPM). This way, the shape of an object can be taken into consideration to produce predictions of surface radiances and temperatures. Deviations in the observed data from predictions of the model can help identify geographically localised features that would otherwise be impossible to find (e.g. low-level conductive heat flow), and allow surface properties to be constrained. On a planetary scale, a topographic thermal model captures a much greater degree of variation in temperatures than simple smooth sphere models, which can lead to more accurate predictions. With an ever-expanding selection of detailed shape-models paired with flyby and orbital thermal observations of asteroids, comets, and moons from missions such as ESA's Rosetta and Juice, as well as NASA's OSIRIS-REx and Europa Clipper, there is an increasing need for intuitive, computationally efficient TPMs that accurately consider the complex topographies of these bodies at high resolutions. 

For example, the icy moons Europa (of Jupiter) and Enceladus (of Saturn) are widely regarded as some of the most promising environments in the Solar System for the potential discovery of extraterrestrial life. A key indicator of their habitability is the flux of heat conducted from subsurface oceans through their icy shells \citep[e.g.][]{2006Spencer, howett2011high}. However, quantifying and interpreting this heat flux using spacecraft-based remote sensing is complicated by factors including diurnally and seasonally varying solar insolation, surface roughness, topographic variation, and heterogeneity in physical properties \citep[e.g.][]{SPENCER1989337,rozitis2011directional}. An essential tool for disentangling these effects is the TPM, which provides a physics-based link between observed radiance and subsurface properties.

Thermophysical models have been widely developed and applied to asteroids \citep[e.g.][]{harris2002asteroids,hanuvs2015thermophysical}. These targets provide ideal test beds for thermal modelling, with extensive literature for comparison and validation. However, many existing TPMs are not publicly available, and few have been designed with extensibility in mind. Extending their application to new environments, such as icy moons, presents several additional challenges:

\begin{enumerate}
    \item Capturing the effects of complex topography, including self-shadowing, radiative self-heating, and mutual scattering.
    \item Incorporating internal or endogenic heating processes \citep[e.g.][]{howett2011high}.
    \item Representing non-Lambertian scattering, which has been shown to be critical on the Moon \citep[e.g.][]{king2020oxford} and may be similarly important on worlds like Europa, where small variations in illumination due to scattering could increase surface temperatures in double ridge structures into a regime of volatile loss \citep{sorli2026thermal}.
    \item Achieving computational efficiency; capturing the small-scale topography of moons makes modelling significantly more demanding than smooth surface models. With high-resolution global topographic data -- such as those soon to be available for Europa -- a model that efficiently captures the effects of resolved small-scale topography across the entire globe will be useful for studying the endogenic heat budget and transport of volatile species.
    \item Extending one-dimensional solvers to account for processes such as sublimation, variation of material properties with depth, and the solid-state greenhouse effect, which is driven by the partial transparency of ice to incident solar radiation and its opacity at thermal infrared wavelengths.
\end{enumerate}

These challenges motivated the development of a new, open-source, Python-based TPM: the Thermophysical Evolution Model for Planetary Environment Surface Temperatures (TEMPEST). TEMPEST was designed to be modular, efficient, and adaptable, with support for a wide range of physical processes, including self-shadowing, radiative self-heating, non-Lambertian scattering, and a flexible structure that allows the user to incorporate alternative one-dimensional solvers to capture effects such as the solid-state greenhouse effect, and endogenic heating.

The Python programming language has historically had a reputation in numerical simulation as being ``fast to write, but slow to run", a view concisely and effectively challenged by \cite{augier2021reducing}. With careful use of optimisation techniques such as vectorisation; as well as tools including Numba (just-in-time compiler) and Joblib (parallelisation across multiple CPUs), \textcolor{revtwo}{TEMPEST achieves efficient computational performance} across a wide range of hardware without compromising readability and usability.

Following the initial development of TEMPEST, the focus shifted to validation and scientific application. The results are found to be consistent with established models, namely, the one-dimensional thermal conduction model, Thermprojrs by \cite{SPENCER1989337}, and the three-dimensional surface energy balance model, the Binary Thermophysical Model (BTPM) by \cite{sorli2025thermophysical}. Validation of the morphology-dependent effects of shadowing, scattering, and mutual heating was done by checking for internal consistency (straightforward for shadowing calculations), and by comparing the results with those of the BTPM. Further validation was performed using empirical data from the Lunar Reconnaissance Orbiter Diviner dataset.

TEMPEST was built to be a shared and collaborative tool. Beyond the applications described in this paper, the model has begun to gain traction in the wider community \citep[e.g.][]{braga2026surface, howes2026thermal}, with early external users already applying it to their own science cases. Ongoing development aims to encourage further adoption and contributions, ensuring that TEMPEST can continue to support planetary science investigations across a wide range of Solar System bodies.


\section{Existing Three-Dimensional Thermophysical Models}
\label{sec2}

The thermal behaviour of airless planetary bodies is governed by their surface and subsurface properties including Bond albedo, \textcolor{revone}{emissivity,} surface roughness, and thermal inertia. Thermal inertia ($\Gamma$) has units of \tiu, and is a measure of a material's resistance to changes in temperature. It is given by $\Gamma = \sqrt{\kappa\,\rho\,c}$ where $\rho$ is density in kg\,m$^{-3}$, $\kappa$ is thermal conductivity in W\,m$^{-1}$\,K$^{-1}$, and $c$ is specific heat capacity in J\,kg$^{-1}$\,K$^{-1}$. Accurate thermal models are essential for interpreting observational data and planning observations. Traditional one-dimensional models have provided valuable insights into the thermal evolution of these bodies by simulating one-dimensional heat conduction. These models play an important role in understanding the diurnal and seasonal temperature variations of planetary surfaces.

However, the complexity of three-dimensional topographic features such as craters, ridges, and boulders necessitates more advanced modelling techniques. Table~\ref{tab:tpms} gives an overview of some of the most widely cited thermophysical models.
Each of the models listed solves a surface energy balance for each facet of a shape-model file, \textcolor{revtwo}{equating the absorbed solar flux to the sum of thermal emission, conduction into the subsurface and, where included, internal heat flux from below (see Equation~\ref{eq:energy_balance} for the form used in TEMPEST).} \textcolor{revone}{Many of the models listed further account for shadowing, light scattering (Lambertian or otherwise), and radiative heating from other visible facets, as well as sub-facet roughness.} For a comprehensive review of the most widely used TPMs as of 2015 see \cite{delbo2015asteroid}. 


\begin{table*}
\caption{Comparison of a representative selection of thermophysical models and their supported capabilities as found in the literature at the time of writing. Note that `shadowing' and `radiative self-heating' are used here in the context of topographic effects, rather than those due to sub-facet scale surface roughness.}
\vspace{-10pt}
\label{tab:tpms}

\begingroup
\scriptsize
\setlength{\tabcolsep}{3pt}
\renewcommand{\arraystretch}{1.15}

\centering
\begin{tabular}{|l|l|l|l|*{7}{c|}}
\rule{0pt}{1cm}%
\textbf{Model Name} &
\textbf{Code availability} &
\textbf{Citation} &
\textbf{Coding language} &
\makebox[0.5cm][l]{\rotatebox{30}{\parbox{2.4cm}{\hspace{0.4pt} \\Shadowing}}} &
\makebox[0.5cm][l]{\rotatebox{30}{\parbox{2.4cm}{\hspace{0.4pt} \\Radiative self-heating}}} &
\makebox[0.5cm][l]{\rotatebox{30}{\parbox{2.4cm}{\hspace{0.4pt} \\Non-Lambertian scattering}}} &
\makebox[0.5cm][l]{\rotatebox{30}{\parbox{2.4cm}{\hspace{0.4pt} \\Roughness}}} &
\makebox[0.5cm][l]{\rotatebox{30}{\parbox{2.4cm}{\hspace{0.4pt} \\Modular 1D solver}}}\\
\hline
TEMPEST & GitHub - duncanLyster/TEMPEST & This work & Python &
Yes & Yes & Yes & Yes & Yes\\
Mocsi & ReadtheDocs - mocsi.readthedocs.io & \cite{schuckart2025modular} & C++ &
Yes & Yes & No & No & Yes\\
Binary TPM & GitHub - ksorli/BinaryThermophysicalModel & \cite{sorli2025thermophysical} & Python &
Yes & Yes & No & No & No\\
Kalast & GitHub - GregoireHENRY/kalast & \cite{henry2024thermal} & Rust (+Python wrapper) &
Yes & Yes & No & No & No\\
Wohlfarth Model & Not open-source & \cite{wohlfarth2023advanced} & Not publicly specified &
Yes & Yes & Yes & Yes & No\\
Asteroid TPM & GitHub - Astroshaper/AsteroidThermoPhysicalModels.jl & \cite{kanamaru2022development} & Julia &
Yes & Yes & No & No & Yes\\
O3DTM & GitHub - tw7044/O3DTM & \cite{king2020oxford} & Matlab &
Yes & No & Yes & No & No\\
VS-TPM & Not open-source & \cite{hanuvs2015thermophysical} & Not publicly specified &
No & No & No & Yes & No\\
KRC & ASU site - krc.mars.asu.edu & \cite{kieffer2013thermal} & FORTRAN &
No & No & No & No & No\\
Mueller (TPMconvex) & GitHub - MigoMueller/TPMconvex & \cite{muller2012surface} & C++ &
No & No & No & Yes & No\\
ATPM & \textcolor{revtwo}{GitHub - SamuelLeeJackson/BinaryTPM} & \cite{rozitis2011directional} & C++, CUDA &
Yes & Yes & No & Yes & No\\
ThermOPS & OCA site - www-n.oca.eu/delbo/ & \cite{delbo2004nature} & C &
No & No & No & Yes & No\\
Lagerros & Not open-source & \cite{lagerros1998thermal} & Not publicly specified &
No & No & No & Yes & No\\
\hline
\end{tabular}
\endgroup

\end{table*}

All of the open-source models listed in Table~\ref{tab:tpms} are well validated and many are staples in the field. While existing three-dimensional models have addressed some of these complexities, there remains a need in the planetary science community for accessible, open-source, models to simulate and interpret thermal data from current and upcoming missions accurately and in high-spatial resolution. There is currently no TPM known to the authors that includes both \textcolor{revone}{look-up table based sub-facet} roughness effects and non-Lambertian scattering (except as a consequence of computationally intensive physically modelled roughness). This is particularly important for regions that are highly shadowed, such as permanently shadowed craters on the Moon, or the neck region of bilobate targets, where both phenomena can have a substantial influence on surface temperatures. 

The aim in developing this open-source Python-based model is to make advanced thermal modelling techniques available to a wider scientific community by providing a comprehensive tool that can account for three-dimensional morphological features including shadowing effects, roughness, and complex scattering behaviour. The result is a temperature map that evolves over time, providing insights into the diurnal cycle and the thermal properties of the target body under various conditions. Further, the user can apply the built-in roughness scheme to create simulated radiance measurements, which can be used for comparison with thermal instruments, reduction of thermal spectra, and mission planning.


\section{Model Overview}
\label{sec3}


\subsection{Outline of the model}
\label{subsec1}

\begin{figure*}
    \centering
    \includegraphics[trim={0 0 0 0},clip, width=0.9\linewidth]{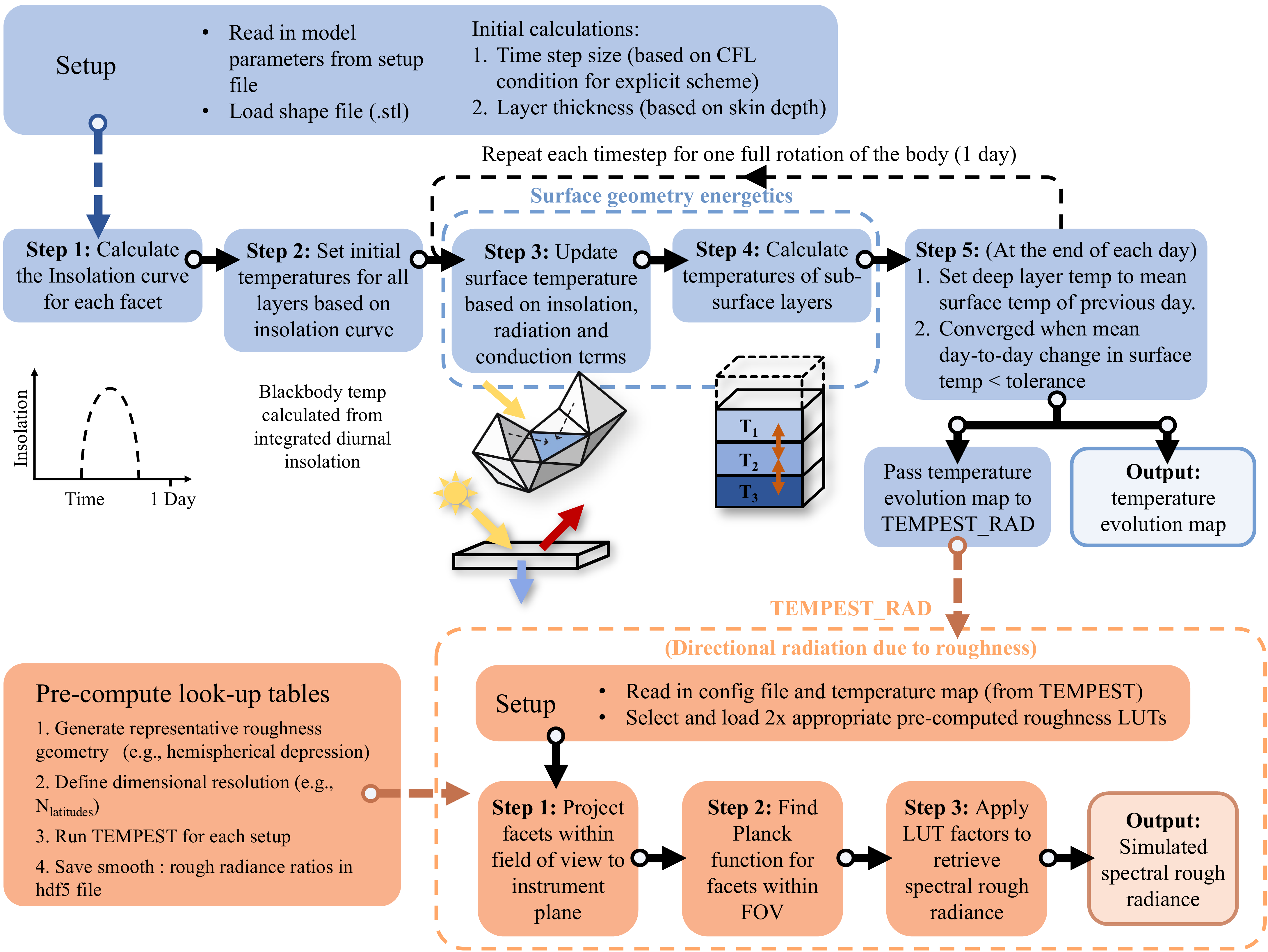}
    \caption{Overview of the logical flow of the TEMPEST model. This roadmap covers the key steps in the `smooth' model calculation of surface temperatures in blue, and the steps taken to retrieve the simulated roughness influenced spectral radiance.}
    \label{fig:code_flowchart}
\end{figure*}

The flowchart in Figure~\ref{fig:code_flowchart} details the workflow behind the model. During the setup stage, the model reads in key parameters from the \texttt{config.yaml} file. These include the physical properties of the target, the path to a shape-model in \texttt{.stl} format, and the target ephemeris and rotation state. When run with the explicit \texttt{tempest\_standard} solver, the time step is selected automatically unless a value is specified by the user. The automatic time step ($\Delta t$) selection is based primarily on the Courant-Friedrichs-Lewy (CFL) stability condition for the one-dimensional finite-difference heat diffusion equation (e.g. \citealt{LeVeque_2002}). For a layer thickness $d$ and thermal diffusivity ($\alpha = \kappa/(\rho c)$), this requires that

\begin{equation}
\frac{\alpha \Delta t}{d^2} \leq \frac{1}{2}.
\label{eq:cfl}
\end{equation}

TEMPEST therefore chooses a time step safely below this limit (multiplying by an additional safety factor of 0.8), then converts this into an integer number of time steps per rotation period. A secondary empirical constraint is also applied to prevent excessively large temperature changes in the surface layer during a single time step. The user may instead prescribe the number of time steps per rotation directly in the configuration file, although the model issues a warning if the chosen value violates the explicit-solver stability criterion.

\textcolor{revone}{The model does not currently support time-dependent variation of sub-solar latitude or heliocentric distance within a model run to capture seasonal heating effects, but the impact of seasonality can be approximated by comparing diurnal temperatures between runs with these set to summer and winter configurations for a given target body. Eclipses can be modelled by manually setting insolation values to zero during a range of timesteps. The default one-dimensional solvers do not allow the user to vary thermophysical properties with depth or temperature, but the modular structure of the code means that, in principle, the model can be paired with other one-dimensional schemes that account for these effects.
}

\subsection{Initialisation and model architecture} 
\label{subsec:Initialisation}

The main functionality of the code is split into 5 main steps as follows:

\textbf{Step 1:} The first step is to determine which facets of the model can `see' each other. This is done by computing a visibility matrix where each entry in the matrix indicates whether one facet is visible to another, allowing calculations involving shadowing and radiative heat transfer to be ignored for facets that are not in one another's line of sight. After visibility is established, TEMPEST estimates geometric view factors using a Monte Carlo ray tracing approach detailed in Subsection~\ref{subsec:viewfactors}. 

\begin{figure}
    \centering
    \includegraphics[trim={0 0 0 0},clip, width=0.9\linewidth]{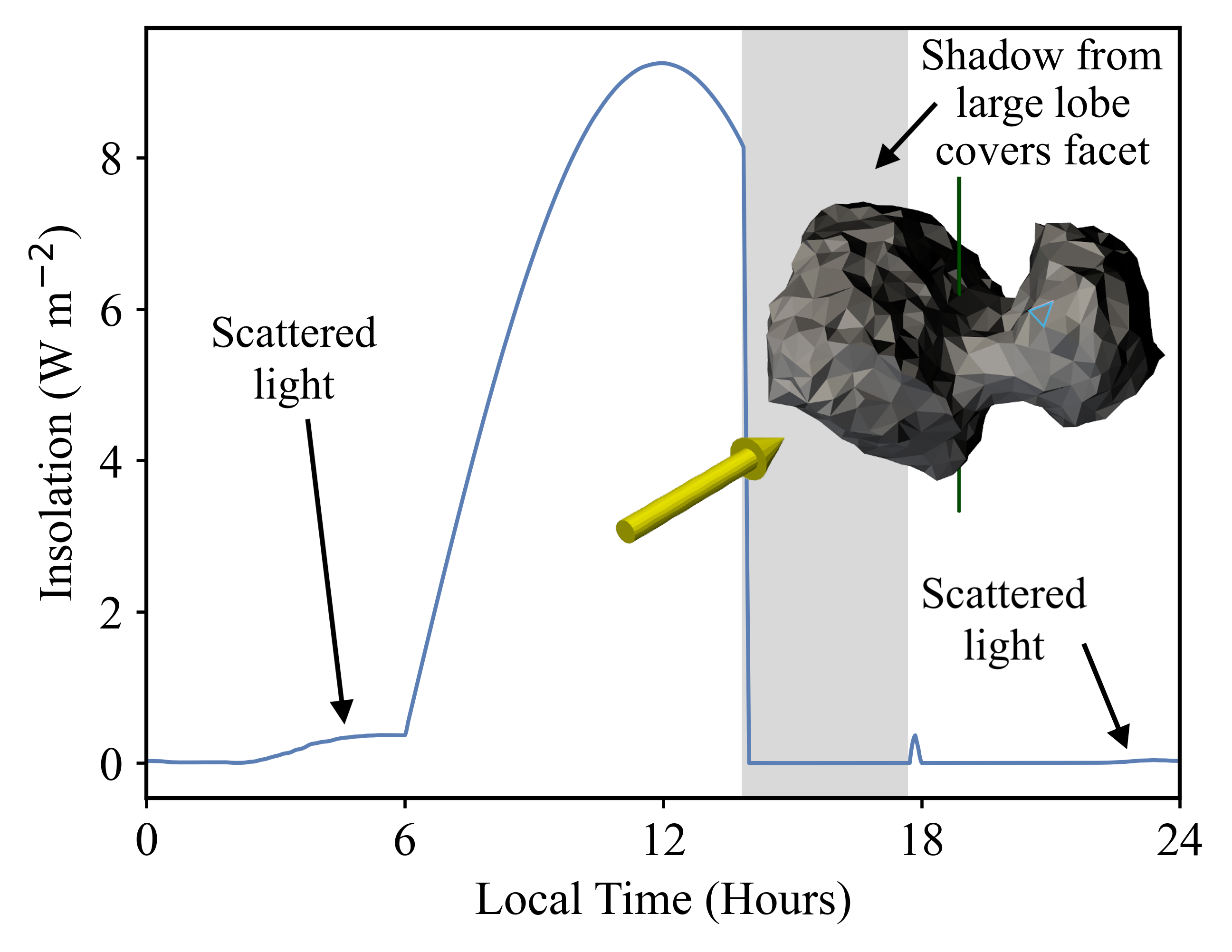}
    \caption{An example insolation curve from a 1,666 facet model of the bilobate comet 67P/Churyumov–Gerasimenko (downsampled from the model first presented by \protect\citealt{jorda2016global}). The main peak at noon is due to direct illumination which is cut off at $\sim$13:00 as the facet moves into shadow \textcolor{revone}{(shown by the grey shaded region)}. The effects of scattered light from other facets can be seen in the 00:00-06:00 region, and the small increase at $\sim$23:00. The axis of rotation and sunlight direction are shown. The selected facet is outlined in blue.}
    \label{fig:insolation}
\end{figure}

The insolation (\(I_{i}\)) for each facet, $i$, over time is calculated as follows:

\begin{equation}
I_{i} = \frac{L_\odot (1 - a) \psi_{i} \cos \theta_{i,z}}{4 \pi r_\odot^2} + I_{\beta}(1 - a),
\label{eq:insolation}
\end{equation}

where \(L_\odot\) is the solar luminosity in W, \(a\) is the Bond albedo, \(\psi_{i}\) is the illumination factor accounting for shadowing (0 or 1; calculated as per subsection~\ref{subsec:insolation}), \(\theta_{i,z}\) is the zenith angle between the facet normal (z) and the sunlight direction, \(r_\odot\) is the distance from the Sun to the target in metres. \(I_{\beta}\) is light scattered from other facets, calculated as follows: 

\begin{equation}
I_\beta = \sum_{j} a\, I_j V_{ij} f_s(\theta_0,\theta_s,\phi_s),
\label{eq:scattered}
\end{equation}

where $I_j$ is the insolation received by facet $j$, $V_{ij}$ is the view factor between facets $i$ and $j$, and $f_s(\theta_0,\theta_s,\phi_s)$ is a directional scattering factor derived from the look-up table described in Subsection~\ref{subsec:scattering}. The sum is taken over all facets $j$ visible to facet $i$. Multiple scattering can be applied by recalculating the $I_\beta$, using the updated insolation curve each time - this is particularly important for heavily shadowed facets on high-albedo targets. Figure~\ref{fig:insolation} shows the effect of scattered light in an insolation curve taken from the model. 

\textbf{Step 2:} Set initial temperatures for all layers based on the insolation curve as a reasonable initial estimate. Each layer of the model is initialised to the blackbody temperature corresponding to the average energy input calculated in Step 1. This assumes that the body is initially in thermal equilibrium with the mean solar input, so the temperature of each facet is set according to the Stefan-Boltzmann law, ensuring a consistent starting point for the simulation.

\textbf{Step 3:} Update surface temperatures based on insolation, radiation, and conduction terms. During this step, the model iteratively calculates the new surface temperature for each facet by considering the balance of incoming solar radiation, thermal radiation emitted from the surface, and conductive heat transfer within the material. The surface temperature \(T_{i,t}\) at time step \(t\) for facet \(i\) is governed by the energy balance of the top layer ($l=0$), which equates the rate of change of its temperature to the net flux of energy into the layer. \textcolor{author}{Taking $\rho c d/2$ as the heat capacity of the surface layer per unit area (the surface node represents the upper half of the first layer), }this balance can be written:

\begin{equation}
\frac{\rho c d}{2} \, \frac{\mathrm{d}T_{l=0}}{\mathrm{d}t} = I_i - \epsilon \sigma T_{i,t-1}^4 + \frac{\kappa}{d}\left(T_{l=1,t-1} - T_{l=0,t-1}\right) + \epsilon^2 \sigma \sum_{j} V_{ij}\, T_{j,t-1}^4 .
\label{eq:energy_balance}
\end{equation}

\textcolor{revone}{where \(\epsilon\) is the emissivity and \(\sigma\) is the Stefan-Boltzmann constant.} The four terms on the right-hand side are (from left to right) the insolation $I_i$, the thermal radiation emitted by the surface, the conductive heat exchange with the first subsurface layer, and the thermal radiation received from other facets. We solve this using an explicit forward difference in time; dividing through by $\rho c d/2$ gives the temperature update, which we group into separate contributions for each process:

\begin{equation}
T_{i,t} = T_{i,t-1} + \left( \Delta T_{i,\odot} + \Delta T_{i,\epsilon} + \Delta T_{i,\kappa} + \Delta T_{i,\gamma} \right),
\end{equation}

where \(T_{i,t-1}\) is the temperature at the previous time step, and the four terms represent total absorbed insolation (\(\Delta T_{i,\odot}\)), radiative cooling (\(\Delta T_{i,\epsilon}\)), conductive heat transfer (\(\Delta T_{i,\kappa}\)), and secondary radiation from surrounding facets (\(\Delta T_{i,\gamma}\)). Each term reflects the energy balance on the surface of the facet, ensuring an accurate representation of the physical processes. $\Delta T$ is used to indicate that the temperature contributions are a finite adjustment applied in each time step as a result of applying the finite difference method to solve a continuous differential equation. The temperature change due to solar radiation absorbed by the facet, $\Delta T_{\odot}$, is calculated as:

\begin{equation}
\Delta T_{\odot} = \frac{I_i \Delta t}{\rho c d / 2},
\label{eq:solar_temp_change}
\end{equation}

where \(I_i\) is the insolation on facet \(i\). Shadowing is contained within $I_i$ through the factor $\psi_i$ (see Subsection~\ref{subsec:insolation}). The temperature change due to thermal radiation emitted by the surface, \(\Delta T_{\epsilon}\), is calculated as:

\begin{equation}
\Delta T_{\epsilon} = - \frac{\epsilon \sigma T_{i,t-1}^4 \Delta t}{\rho c d / 2},
\end{equation}

where \(T_{i,t-1}\) is the surface temperature at the previous time step. The temperature change of the surface layer \(l=0\) due to conductive heat exchange with the subsurface layer, \(\Delta T_{\kappa}\), is calculated as:

\begin{equation}
\Delta T_{\kappa} = \frac{2 \alpha \Delta t}{d^2} \left(T_{l=1,t-1} - T_{l=0,t-1}\right),
\end{equation}

where \(T_{l=1,t-1}\) is the temperature of the first subsurface layer ($l=1$) at the previous time step ($t-1$). Lastly, the temperature change due to secondary radiation from other facets, \(\Delta T_{\gamma}\), is given by:

\begin{equation}
\Delta T_{\gamma} = \frac{\epsilon^2 \sigma \Delta t}{\rho c d  / 2} \sum_{j} V_{\gamma, ij} T_{j,t-1}^4,
\label{eq:self_heating}
\end{equation}

where the sum is taken over all facets (j) visible to facet (i). Equation~\ref{eq:self_heating} assumes isotropic (Lambertian) thermal emission. Optionally, the user may apply a static thermal emission phase function in direct analogy to the directional scattering factor of Subsection~\ref{subsec:scattering}, multiplying each $V_{ij}$ by a coefficient that depends on the emission angle from facet $j$ to facet $i$ to create a new set of static thermal view factors $V_{\gamma, ij}$. \textcolor{revone}{For an idealised smooth surface} this case is simpler than the scattering one, as thermal emission has no incident ray and the coefficient therefore depends on emission angle alone (see Subsection~\ref{subsec:viewfactors}). In the Lambertian case this coefficient is unity at all emission angles and the thermal view factors reduce to $V_{ij}$. \textcolor{revone}{For a real surface with unresolved roughness, however, sunlit slopes are warmer than shaded ones, so the directional emission also depends on illumination geometry and an emission-angle-only coefficient is an approximation. Where this effect matters, roughness is better captured by resolving it explicitly in the shape model.}

\textbf{Step 4:} Calculate temperatures of subsurface layers. The temperature of each subsurface layer is updated by solving the heat conduction equation, using the finite difference method to approximate the temperature gradient between layers. The finite difference method allows the model to iteratively calculate the temperature changes within the subsurface layers by considering the heat flow between adjacent layers.

The temperature for a subsurface layer \(l\) at time step \(t\), denoted as \(T_{l,t}\), is updated according to:

\begin{equation}
T_{l,t} = T_{l,t-1} + \frac{\alpha \Delta t}{d^2} \left(T_{l+1,t-1} - 2T_{l,t-1} + T_{l-1,t-1}\right).
\end{equation}

This equation considers the heat conducted from the layer above and below, ensuring that the temperature profile within the subsurface is accurately captured over time.

The finite difference method is employed because it discretises the continuous heat conduction equation, allowing computation of the thermal evolution of the subsurface layers over time. By iterating this calculation across all layers and time steps, the model simulates the temperature distribution within the subsurface. This approach accounts for the complex interactions between surface and internal heat flows, which ultimately influence the thermal inertia, a quantity which can be inferred from observations. 

\textcolor{revtwo}{\textbf{Semi-implicit solver:} Steps 3 and 4 describe the default explicit solver (\texttt{tempest\_standard}). Alternatively, the user can select the semi-implicit solver (\texttt{tempest\_implicit}) in the \texttt{config.yaml} file. This solves the same energy balance, but rather than updating each layer using temperatures from the previous time step, it solves for the temperatures of all layers of a facet at the new time step simultaneously. Heat conduction between subsurface layers is calculated using the Crank-Nicolson method, which takes the average of the conductive fluxes at the start and end of each time step. As in the explicit solver, the surface layer has heat capacity $\rho c d/2$, and its energy balance is solved fully implicitly. To keep the set of equations linear, the thermal emission term is linearised about the surface temperature at the previous time step:}

\begin{equation}
\color{revtwo} T_{i,t}^4 \approx 4T_{i,t-1}^3\,T_{i,t} - 3T_{i,t-1}^4.
\end{equation}

\textcolor{revtwo}{Insolation is averaged over the start and end of each time step, and radiation from other facets is taken from the previous time step, as in the explicit solver. Since each layer is linked only to the layers directly above and below it, the resulting set of equations for each facet forms a tridiagonal matrix, which is solved efficiently using the Thomas algorithm, a simplified form of Gaussian elimination \citep[e.g.][]{alma990194379310107026}. A zero-flux condition is applied at the base of the layer grid, and the deep-layer reset described in Step 5 is not used. The main advantage of this approach is that it is unconditionally stable, so it is not bound by the time step limit of Equation~\ref{eq:cfl} and can be run with fewer time steps per rotation (see Subsection~\ref{subsec:benchmarking}). The disadvantage is that abrupt changes in forcing, such as a facet entering or leaving shadow, can cause short-lived oscillations in the solution (see Subsection~\ref{subsec:internal_validation}).}

\textbf{Step 5:} At the end of each day, the model checks for convergence using the per-facet change in surface temperature between successive days, $\Delta T_i = T_{i,n} - T_{i,n-1}$. By default, the simulation is considered converged when the mean of these changes across all $N$ facets, $\Delta T_{mean} = \frac{1}{N}\sum_i \lvert \Delta T_i \rvert$, falls below a predefined target. Alternatively, the user may adopt a stricter criterion based on the largest change of any individual facet, $\Delta T_{max} = \max_i \lvert \Delta T_i \rvert$. This is useful for slowly converging regions such as permanently shadowed regions, where the mean change can fall below the target while a few cold facets are still evolving. If the chosen criterion is not yet met, the deep layer temperature, $T_{\rm{deep}}$, is set to the radiative mean surface temperature of the previous day, $\left(\langle T^4 \rangle\right)^{1/4}$, and the cycle repeats until convergence is achieved.


\subsection{Insolation and shadowing}
\label{subsec:insolation}

Insolation is calculated geometrically for each facet according to Equation~\ref{eq:insolation}. The shadowing factor $\psi_i$, is determined using a ray-tracing approach and takes a value of 0 or 1. A single ray is cast from \textcolor{revtwo}{the centre of} the subject facet in the direction of the Sun, the model then checks for intersections between this ray and opposing facets using a vectorised version of the M{\"o}ller-Trumbore algorithm \citep{moller2005fast}. 

A number of optimisations are used to speed this process up, first, any facets below the horizon of the subject facet are removed from the set of test facets using their centroid position. Second, any facets that face away from the target facet are removed. The remaining set of facets is passed to the vectorised triangle intersection algorithm. 

Insolation conditions are periodic for principal-axis rotators, repeating exactly for each rotation. This means insolation can be calculated just once for the model, and it can then be run for as many full rotations as are required to reach convergence. The same applies to shadowing and scattering of light. 

\begin{figure*}
    \centering
    \includegraphics[trim={0 0 0 0},clip, width=\linewidth]{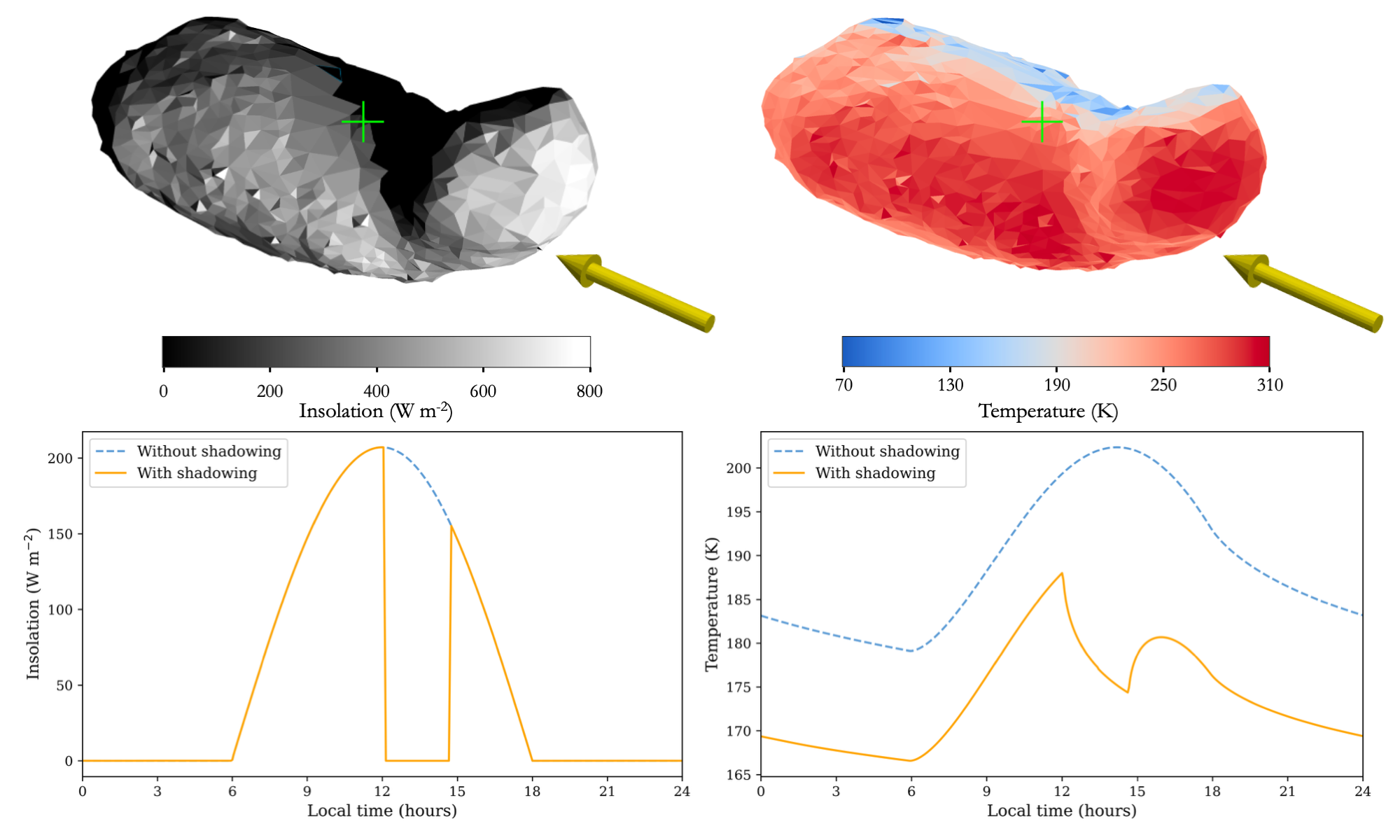}
    \caption{Upper left: Insolation map of a 3,144-facet shape-model of 25143 Itokawa showing the neck region (collum) and part of the large lobe shadowed by the smaller lobe on the right. Upper right: A temperature map showing the TEMPEST result including shadowing. The yellow arrow shows the sunlight direction\textcolor{revone}{, and the green cross shows the location of the facet whose insolation and temperature are plotted in the lower panels.} Lower left: Received insolation of selected facet with (solid) and without (dashed) shadowing. Lower right: Temperature of selected facet with (solid) and without (dashed) shadowing. The rotation state of the body is expressed in 24-hour local time, where a local time of 12:00 corresponds to the minimum solar zenith angle.}
    \label{fig:shadowing}
\end{figure*}

Partial shadowing can be approximated by optionally smoothing the binary illumination state in time after the geometric shadowing calculation. This spreads abrupt changes in insolation over a small number of time steps, during shadow ingress and egress. Although this is not a true penumbra calculation, it is useful when using the semi-implicit solver, as it reduces sharp discontinuities in forcing and helps maintain stable numerical behaviour. The inclusion of shadowing effects increases the accuracy of temperature predictions in areas with significant topographic variation (Figure~\ref{fig:shadowing} - top left). Figure~\ref{fig:shadowing} (top right) shows surface temperatures generated by this model, highlighting the temperature distribution across the surface. These maps can be used to better understand the thermal evolution of the target bodies, variation in their surface properties, and for planning mission operations. 


\subsection{Determination of view factors}
\label{subsec:viewfactors}

At the facet level, calculation of view factors is done using ray-tracing. This approach was chosen to keep overheads manageable on high resolution models with a large degree of concavity. Sampling a representative subset of the visible opposing terrain is less computationally intensive than evaluating every possible facet-facet interaction, and still leads to accurate calculation of scattering and radiative self-heating contributions. First, the full set of facets in the shape-model is reduced based on their preliminary visibility. Specifically, for a given test facet, $i$, facets that face away from it (or that it faces away from) are first removed from the set of target facets. Mathematically, a target facet $j$ is removed if:

\begin{equation}
    (\mathbf{r}_j - \mathbf{r}_i) \cdot \mathbf{\hat{n}}_i \leq 0 \quad \text{or} \quad (\mathbf{r}_i - \mathbf{r}_j) \cdot \mathbf{\hat{n}}_j \leq 0,
\end{equation}

where $\mathbf{r}$ and $\mathbf{\hat{n}}$ represent the centroid position vectors and outward-pointing unit normal vectors of the respective facets. A ray is then created by subtracting the centroid of facet $i$ from that of $j$. This ray is fired from the centroid of $i$ towards $j$ and compared to the remaining set of facets to check for an intersection. If an intersection occurs, the target facet $j$ is removed from the set. This way, occluded facets are removed. Using centroids means partial occlusion of facets is not considered, meaning illuminated regions near the edge of the facet are missed -- particularly in rough or spiky terrains. The impact of this on predicted surface temperatures can, however, be reduced by using higher-resolution shape-models or even by subdividing facets in a shape-model.

To go from a set of visible facets to a set of view factors, TEMPEST estimates the geometric view factor \(V_{ij}\) by Monte Carlo ray tracing. A user-input number of rays, \(n\), \textcolor{revone}{are fired from random points within the triangle of the subject facet \(i\) in directions drawn from a cosine-weighted (Lambertian) distribution over the hemisphere, so that the fraction of rays striking facet j is an unbiased estimate of the diffuse view factor. These} rays are checked for intersections with the set of visible facets using the same intersection algorithm as in Subsection~\ref{subsec:insolation}. \(n_j\) of \(n\) rays strike each target facet, \(j\), and the estimated view factor becomes
\begin{equation}
V_{ij} = \frac{n_j}{n}.
\end{equation}

Mutual heating, or radiative self-heating, is an important effect both above and below the spatial-resolution scale of a shape-model. This effect is considered in TEMPEST at spatial-scales greater than the thermal skin depth, i.e., where lateral conduction of heat can be considered negligible - see Equation~\ref{eq:self_heating}. The user has the option of including a static thermal emission phase function (EPF) as a look-up table (LUT). The LUT takes the form of a string of multiplication factors corresponding to emission angles 0-90$^{\circ}$. The `thermal view factor' is found by multiplying the view factor by the corresponding coefficient for emission angle of the subject facet to the test facet.


\subsection{Non-Lambertian scattering of light}
\label{subsec:scattering}
Non-Lambertian scattering of light due to small-scale surface structure has been well documented on planetary bodies since the Apollo era \citep{hapke1963photometric}. Insolation scattering effects caused by surface geometry below the resolution scale of the shape-model are governed by the shadow hiding opposition effect (SHOE) and the coherent backscatter opposition effect (CBOE) \citep{hapke1998opposition}, and as \cite{king2020oxford} showed it can have a significant impact on surface temperatures, particularly in permanently shadowed regions (PSRs). Despite this, non-Lambertian scattering is rarely included in TPMs. Modelling it accurately means tracking the direction of scattered light rather than its magnitude alone, leading to additional complexity in the model architecture. As Table~\ref{tab:tpms} shows, the only currently available open-source model that accounts for any type of non-Lambertian scattering is the O3DTM of \cite{king2020oxford}. 

\begin{figure}
    \centering
    \includegraphics[trim={0 0 0 0},clip, width=\linewidth]{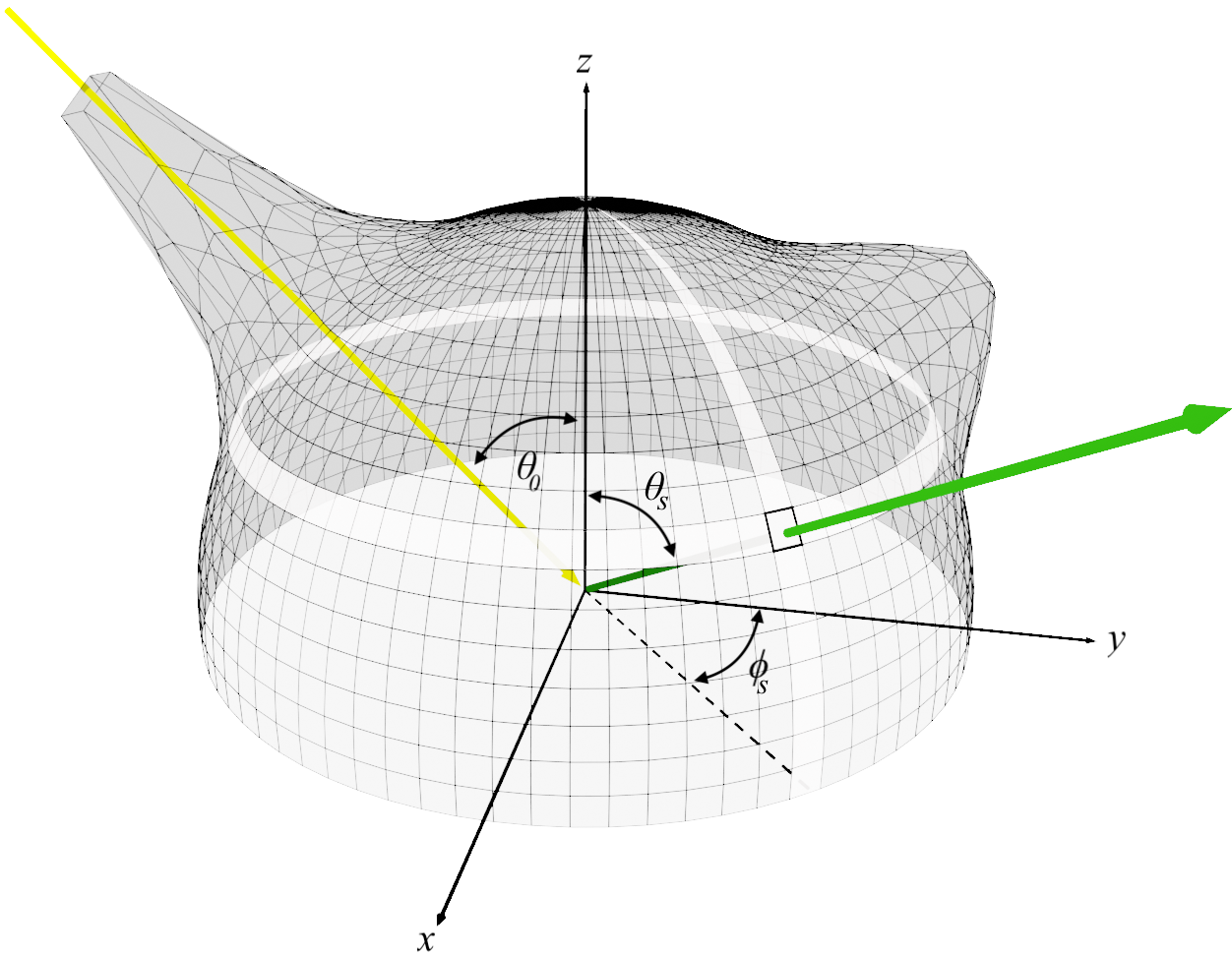}
    \caption{Light scattering look-up table (LUT) geometry in TEMPEST showing incident beam (yellow) and scattered light (green). The three key angles are shown: the bidirectional reflectance distribution function (BRDF) zenith angle of the incident ray $\theta_0$, zenith angle of the scattered ray $\theta_s$, and azimuthal angle $\phi_s$ of the scattered ray. The mesh surface is a visualisation of the method used to discretise incident and emitted rays using a dome. This LUT is defined such that for Lambertian scattering every value of the directional scattering factor, $f_s(\theta_0,\theta_s,\phi_s)$ is unity, so the BRDF must be normalised by fixing the volume enclosed by the scattering surface to 2\(\pi\)/3, the volume of a unit hemisphere. As the shape of the mesh surface shows, this example exhibits a sharp backscattering peak and a smaller, broader forward scattering peak.}
    \label{fig:scattering_lut}
\end{figure}

While a Lambertian assumption is sufficient for many applications, including non-Lambertian scattering extends the model's reach to bodies and regimes where it becomes important, for example, strongly non-Lambertian surfaces, and high latitude shadowed regions such as the lunar PSRs, where scattered light makes a first-order contribution to the local energy budget. Capturing these effects was considered a priority during the development of TEMPEST as a general-purpose model, and is achieved using a framework capable of capturing any arbitrary bidirectional reflectance distribution function (BRDF). This takes the form of a LUT made from a grid of points distributed across the surface of a hemisphere for each incidence angle (Figure~\ref{fig:scattering_lut}). 

For a given illumination and scattering geometry the directional scattering factor $f_s(\theta_0,\theta_s,\phi_s)$ is obtained by trilinear interpolation over the LUT where $\theta_0$ is the zenith angle of the incident ray, and $\theta_s$ and $\phi_s$ are the zenith and azimuthal angles of the scattered ray, respectively. The LUT is normalised to enclose the volume of a unit hemisphere $(2\pi/3)$, meaning $f_s$ redistributes the scattered energy in angle without changing its hemispherical integral, and reduces to unity at all angles in the case of Lambertian scattering. It enters the scattered-light term (Equation~\ref{eq:scattered}) as a multiplier on the Lambertian flux. The user defines the angular resolution for the LUT, and could increase it to capture a steep opposition effect. 

This is a flexible framework as it allows the user to either use the simple Lambertian case, or to create a LUT that captures different scattering behaviour. For example, using lab-derived scattering parameters e.g. from Lunar regolith \citep{curtis2022measuring}, or by using a Hapke BRDF generator (e.g. \citealt{hapke2012theory}), matching Hapke parameters derived from remote sensing, or comparison with similar bodies. 

\subsection{Roughness methodology in TEMPEST}
\label{sec:spherical_depressions}

Surface roughness is a fundamental feature of all planetary bodies, and its influence on thermal processes has long been recognised (e.g. \citealt{spencer1990rough}, \citealt{rozitis2011directional}, \citealt{curtis2022measuring}). Variations in surface slope at sub-facet scales affect both the absorption and emission of thermal energy by introducing shadowing, mutual heating, and directional emissivity effects. These effects can significantly impact radiance measurements and therefore the predicted surface temperatures and thermal inertia values inferred from them. While a smooth-surface model is often sufficient for first-order interpretations (e.g. predicting maximum diurnal temperature), the inclusion of roughness is necessary to accurately reproduce observations, particularly at extreme observing geometries or high phase angles.

A widely adopted and physically motivated method for incorporating surface roughness is to represent unresolved roughness as ensembles of `hemispherical craters' or spherical-depressions\footnote{Here, `spherical-depressions' is used throughout to avoid confusion with resolved impact craters - surface terrain features that may be resolved in a shape-model.} embedded within the surface (e.g. \citealt{lagerros1998thermal}, \citealt{rozitis2011directional}). These geometries are are effective due to their simple parametrisation and have been shown to reproduce key observational signatures, such as thermal‑infrared beaming, particularly under extreme illumination and viewing geometries \textcolor{revtwo}{(\citealt{rozitis2011directional}, \citealt{davidsson2015interpretation}).}

Here we describe the implementation of surface roughness in the TEMPEST model and assess its impact on radiance predictions. The approach builds on established roughness treatments, while introducing adaptations to improve their computational efficiency and maintaining compatibility with the facet structure used in TEMPEST. The following sections outline the modelling framework, validation steps, and initial tests, and an overview of the process is given in Figure~\ref{fig:roughness_flow}.

Simply replacing every facet with a spherical-depression composed of hundreds of sub-facets would be both computationally prohibitive and geometrically inconsistent, since \texttt{.stl} shape-models use triangular facets that cannot be directly reconciled with spherical geometry. To address this, TEMPEST first calculates surface temperatures using a smooth surface model. It then uses a precalculated directional radiative beaming LUT to calculate the wavelength dependent directional flux modification required for each facet based on viewing geometry. The LUT is created using the following steps: 

\begin{figure}
    \centering
    \includegraphics[trim={0 0 0 0},clip, width=\linewidth]{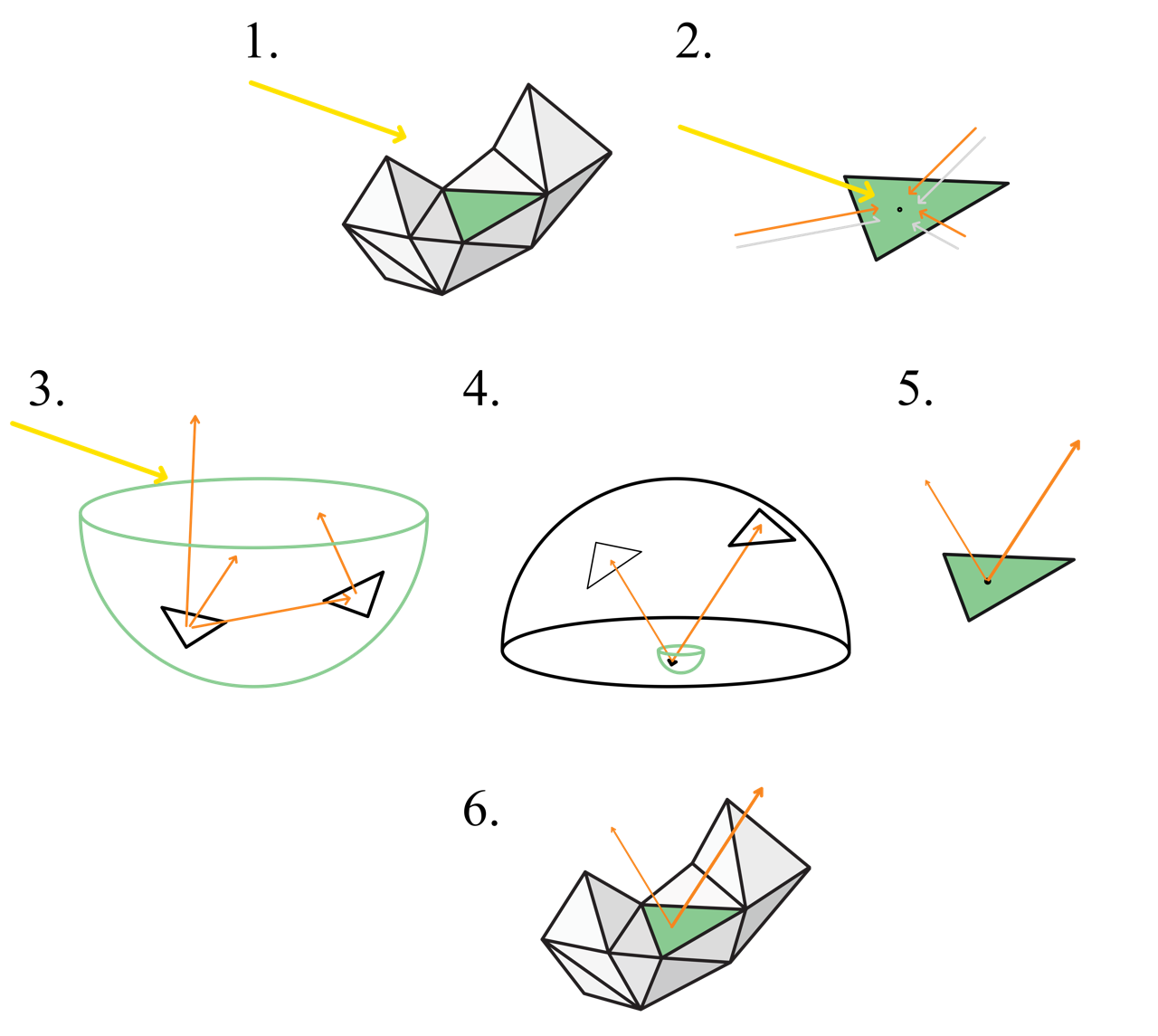}
    \caption{Diagram showing the key steps in the TEMPEST roughness calculations. In steps 1 and 2, the temperature of the highlighted facet is calculated based on the direct insolation (yellow arrow), scattered light (grey arrows), and radiation emitted by other facets (orange arrows), as well as the temperatures of subsurface layers in the previous time step. In step 3 sub-facet temperatures are calculated based on sphere section surface depression geometry, then in step 4 these are used to determine discretised directional spectral emission of radiation from the facet. Steps 3 and 4 can be precalculated and referenced as a look-up table, speeding up model operation. Finally, in steps 5 and 6, directional spectral radiance is calculated for each facet, then combined into a single simulated observation.}
    \label{fig:roughness_flow}
\end{figure}

\noindent\textbf{Step 1 - Parent geometry energetics:} The input shape-model for the target is treated as the ``parent'' geometry, consisting of triangular facets. Energy received by each facet from insolation, accounting for shadowing, plus scattered light and mutual radiative heating is first accumulated in this parent space.

\noindent\textbf{Step 2 - Surface temperature solution:} The surface temperature, $T_i$, of each parent facet is calculated at each time step using the accumulated energy terms and the conductive heat exchange with the subsurface layers. This produces the smooth-surface temperature solution for the full rotation of the body.

\noindent\textbf{Step 3 - Intra-facet energetics:} A representative roughness kernel (a spherical-depression in this case) is then used to calculate the redistribution of energy within a rough facet. \textcolor{revone}{The shape of the depression is set by its opening angle, the half-angle subtended at the centre of the sphere by the depression's rim (0$^\circ$ gives a flat surface, 90$^\circ$ a hemispherical bowl, and larger values a deeper cavity with an overhanging rim). A separate LUT is generated for each opening angle; by default a hemispherical (90$^\circ$) kernel is used.} For each relevant latitude, local time, and \textcolor{revone}{direct solar} illumination geometry, the sub-facets within the depression receive direct insolation, scattered light, and mutual radiative heating from neighbouring sub-facets. \textcolor{revone}{The kernel is forced by direct sunlight alone with an unobstructed horizon, meaning shadowing by resolved terrain, scattered sunlight, and thermal emission arriving from outside the parent facet are not included in the LUT. These terms contribute to the smooth-facet temperature $T_i$ (Steps 1 and 2) but not to the directional radiance factors, so the roughness treatment is an approximation that captures beaming driven by direct solar illumination only (see Subsection~\ref{subsec:roughness_discussion}).} Solving the thermal balance across the kernel gives a set of sub-facet temperatures that describe the unresolved temperature distribution associated with the parent facet. The magnitude of the temperature contrast across the kernel is governed by the dimensionless thermal parameter, $\Theta$, first introduced by \cite{SPENCER1989337}, defined as
\begin{equation}
    \Theta = \frac{\Gamma \sqrt{\omega}}{\epsilon \sigma T_{\text{SS}}^3},
\end{equation}
where $\Gamma$ is the thermal inertia, $\omega = 2\pi/P$ is the angular rotation frequency in rad~$\rm{s}^{-1}$, $\epsilon$ is the surface emissivity, $\sigma$ is the Stefan-Boltzmann constant, and $T_{\text{SS}}$ is the instantaneous sub-solar temperature for a smooth, flat surface. A low $\Theta$ corresponds to large diurnal temperature swings and strong contrast within the cavity; a high $\Theta$ smooths the contrast out. The kernel is solved across a range of $\Theta$, so the radiance factors that follow can be tabulated against it.

\begin{figure}
  \centering
  \includegraphics[trim={0 0 0 0},clip,width=\linewidth]{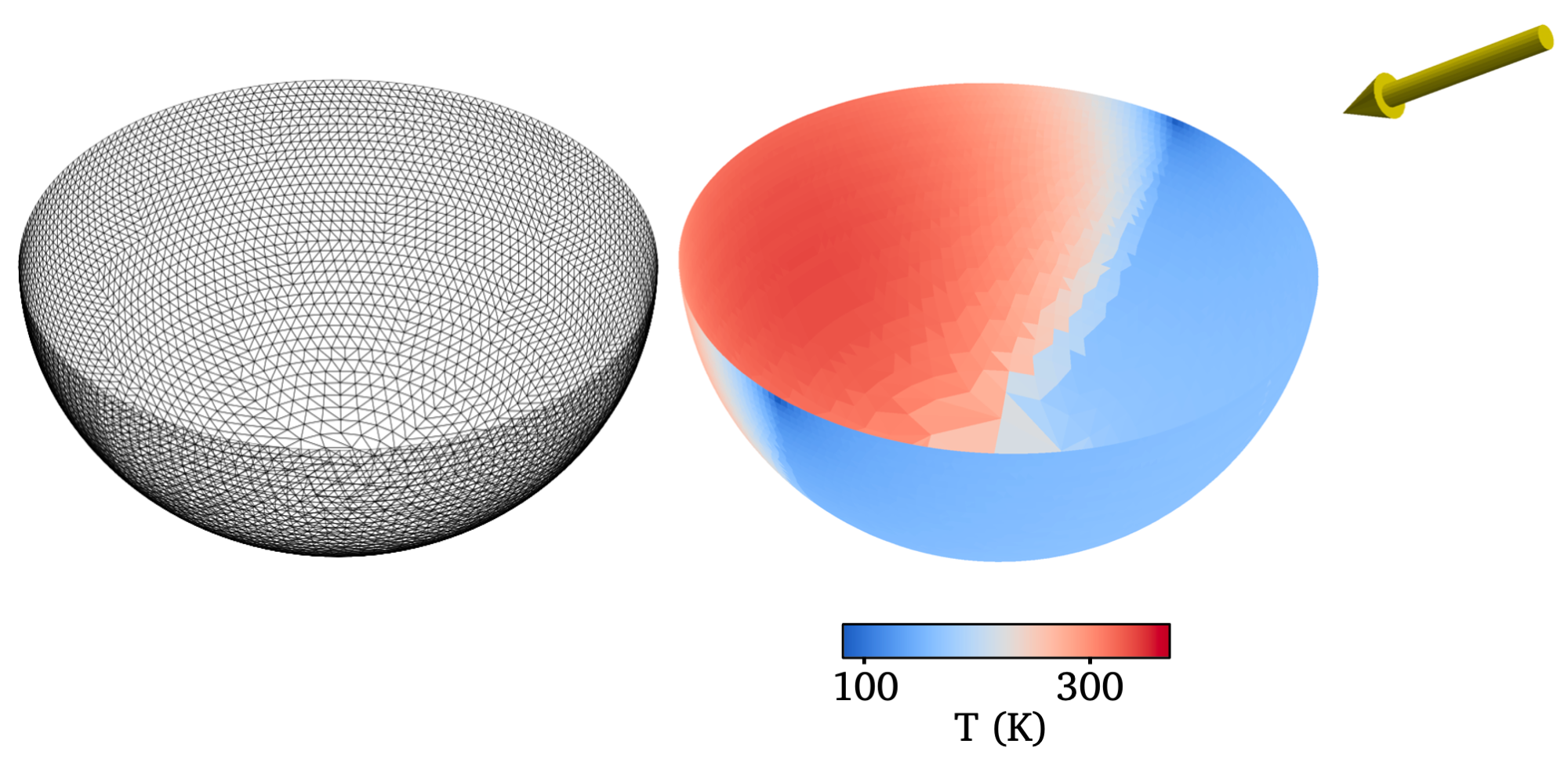}
  \caption{Temperatures of sub-facets within the spherical-depression during look-up table generation, showing an example of the distribution of absorbed energy at the sub-facet level. The yellow arrow indicates the illumination direction.}
  \label{fig:roughness_kernel}
\end{figure}

\noindent\textbf{Step 4 - Directional radiance factors:} The spectral radiance from the roughness kernel (Figure~\ref{fig:roughness_kernel}) is calculated for a discretised set of viewing directions, defined by azimuth and elevation relative to the parent-facet normal. For each wavelength bin and viewing direction, the emitted radiance from all visible sub-facets is combined and compared with the equivalent smooth-facet radiance to give the directional radiance factor
\begin{equation}
    R_\lambda(\theta_v, \phi_v; \Theta) = \frac{L_\lambda^{\mathrm{rough}}(\theta_v, \phi_v)}{B_\lambda(T_{\mathrm{ref}})},
    \label{eq:radiance_factor}
\end{equation}
where $L_\lambda^{\mathrm{rough}}$ is the projected-area-weighted radiance of the rough kernel in the viewing direction $(\theta_v, \phi_v)$, $B_\lambda$ is the Planck function, \textcolor{revtwo}{and $T_{\mathrm{ref}}$ is the temperature of an isolated flat facet with the same thermal parameter, latitude and local time, solved alongside the kernel as a smooth reference.} Every sub-facet shares the same emissivity, so $\epsilon$ cancels between the numerator and denominator, leaving $R_\lambda$ independent of emissivity and acting as a purely geometric thermal redistribution factor. These factors are stored in a roughness LUT.

\noindent\textbf{Step 5 - Facet radiance look-up:} The LUT is queried for each facet using the facet latitude, local time, observer direction, and thermal parameter, $\Theta$, at the relevant time step. The directional spectral radiance of a rough facet toward the observer is then
\begin{equation}
    L_\lambda^{\mathrm{obs}} = \left[(1-f) + f\,R_\lambda(\theta_v, \phi_v; \Theta)\right] \epsilon\, B_\lambda(T_i),
    \label{eq:rough_radiance}
\end{equation}
where $T_i$ is the smooth-facet temperature from Step 2, $R_\lambda$ is interpolated from the LUT for the facet's latitude, local time, $\Theta$, and viewing geometry, \textcolor{revone}{and $f$ is the fractional coverage of the facet by spherical-depressions, with the remainder treated as smooth. The degree of roughness is controlled by $f$ and by the opening angle used to generate the LUT. Depression depth and coverage both contribute to the effective RMS slope of the surface, so following \citet{rozitis2011directional}, by default the opening angle is held fixed at 90$^\circ$ with roughness varied through $f$.}

\noindent\textbf{Step 6 - Full FOV radiance integration:} The directional spectral radiance $L_\lambda^{\mathrm{obs}}$ from all facets within the instrument field of view is then integrated according to their projected visible areas, with facets hidden from the observer excluded. This produces the simulated observation for the specified wavelength range, observing geometry, and model time step. 

\noindent\textbf{Notes:} The number of spectral bins, sub-facets, viewing-direction bins, and latitude and local-time samples used in the LUT are configurable depending on the accuracy and computational cost required for the use case at hand.

\subsection{Discussion of the roughness treatment}
\label{subsec:roughness_discussion}

Decoupling the smooth-surface temperatures from the unresolved roughness is a widely accepted technique (e.g., \citealt{rozitis2011directional}) that makes the computationally expensive problem of thermal beaming tractable. \textcolor{revtwo}{The thermal balance of the kernel includes direct insolation, scattering and radiative self-heating between sub-facets, so over a full rotation the energy it emits through its aperture equals the energy it absorbs (Subsection~\ref{subsec:internal_validation}). The directional radiance factors are taken directly from this solution. At an individual time step, the total power emitted by the kernel over all viewing directions can therefore differ from that of the smooth reference facet, since the cavity stores and releases heat differently. It remains warmer than the smooth facet at night and lags it in the morning. These departures form part of the diurnal beaming signal.}

\textcolor{revtwo}{The radiance factors also capture two effects of roughness on the total emitted power. The first is the non-linearity of thermal emission. A rough surface spanning a range of sub-facet temperatures does not emit as a single surface at the mean temperature would ($\langle T^4 \rangle \neq \langle T \rangle^4$), which raises both its total emitted power and its emission at short wavelengths. The second is cavity trapping. Light scattered within the depression can be absorbed by other sub-facets, so the kernel absorbs more sunlight than the smooth reference facet, and over a rotation a hemispherical kernel emits a few per cent more energy than the smooth reference. As a consequence, the Bond albedo supplied by the user (Equation~\ref{eq:insolation}) should describe an ideally flat surface element rather than the facet as a whole. Unlike the single-scattering albedo, which describes a single grain-level interaction, the Bond albedo integrates over all scattering events. An observationally-derived value therefore already encodes the multi-bounce absorption losses within sub-pixel cavities - a reduction that \citet{spencer1990rough} models explicitly and shows to be dependent on morphology at a scale smaller than that resolved by observations. Retaining the enhanced absorption of the kernel on top of such an albedo would therefore count this effect twice. The spatial redistribution of absorbed energy within the depression is modelled by the sub-facet shape kernel. The accuracy of the resulting temperature distribution therefore depends on how well the spherical section approximates the true surface morphology. Alternative kernel geometries, such as fractal surfaces, may be better suited to specific surface types \citep[e.g.,][]{davidsson2015interpretation, maclennan2024thermal}.}

\textcolor{revtwo}{The decoupled approach has two limitations. The first is that roughness does not feed back into the parent-facet energy balance. The temperature $T_i$, and the scattered light and thermal emission exchanged between parent facets (Equations~\ref{eq:scattered} and \ref{eq:self_heating}), are those of a smooth surface.}

\textcolor{revone}{The second is that the LUT is generated for an isolated facet under direct sunlight, so it does not account for the influence of resolved topography on the illumination of the kernel. This includes shadowing by resolved surface features, as well as scattered sunlight and thermal emission from surrounding terrain. Indirect illumination arrives from a wide range of directions and so tends to heat the depression more uniformly than direct sunlight. Neglecting it will therefore tend to overestimate the directional effects of the beaming correction for facets where indirect illumination makes up a large part of the energy budget, such as those in shadow or within deep concavities.} While it is theoretically possible to account for this by adding parameters for dawn/dusk shadowing or local occultation to the LUT inputs, each new parameter increases the size of the table exponentially. The memory and computational requirements scale as $O(m^N)$ (where $m$ is the number of bins per parameter and $N$ is the number of parameter dimensions), rapidly leading to prohibitive memory constraints. An alternative would be to generate a bespoke LUT for each specific shape-model and illumination geometry, passing the exact insolation conditions to the LUT generator. While this would ensure a perfect match for each facet at each time step and improve directional beaming accuracy, it negates the primary computational advantages of using a generalised, precalculated LUT. As a consequence, the limitations of this roughness approach should be carefully considered when using TEMPEST to simulate bodies with a high degree of macroscopic concavity, or slow rotators with a low thermal parameter where temperature gradients are most extreme.


\section{Validation \& Performance Benchmarking}
\label{sec:Validation}


\subsection{Internal validation}
\label{subsec:internal_validation}
To validate the implementation of surface roughness and the wider TEMPEST framework, a series of internal consistency checks and limiting-case tests were performed. 

The first step was to verify the global energy balance of the model. The Thermprojrs model, on which the core one-dimensional solver in TEMPEST is based, uses strict energy conservation to continuously update the temperature of the deepest subsurface layer. In contrast, TEMPEST applies a boundary condition where the deep layer temperature is set to the radiative mean of the surface layer temperature over the preceding day, $\left(\langle T^4 \rangle\right)^{1/4}$ (Subsection~\ref{subsec:Initialisation}). This approach accelerates convergence to a stable steady-state solution by approximately an order of magnitude. The model does not force a continuous balance to drive the deep-layer thermal solver, meaning global energy conservation can instead be monitored as an independent diagnostic. At equilibrium, the energy absorbed by each facet over one full rotation must equal the energy it emits; any imbalance would imply that energy is being non-physically created or destroyed, and hence a mathematical error in the implementation. We performed this check during development and confirmed that the core one-dimensional solver conserves energy across a full rotation. The solver has not since been modified, so subsequent development has not affected the energy budget of the 1D model.

\textcolor{revtwo}{A second check tests energy conservation within the roughness LUT. Integrating $R_\lambda B_\lambda(T_{\rm ref})$ over all viewing directions, wavelengths and one full rotation gives the total energy emitted by the kernel, which must equal the energy it absorbs over the same period. For the default hemispherical kernel the two agree to within a few per cent across all latitudes and values of $\Theta$.} The roughness implementation is further verified using the limiting case when the spherical-depression \textcolor{revone}{opening} angle is set to zero degrees (essentially a flat disc); \textcolor{revtwo}{the kernel solution should then reproduce the smooth-facet temperatures, giving $R_\lambda = 1$ and radiances identical to a model where the smooth shape-model is used without any sub-facet geometric roughness modifications.} TEMPEST also passes this test.

Finally, internal testing has highlighted specific operational limits to the semi-implicit thermal solver, particularly concerning complex macroscopic topographies. Topographical shadowing on bodies with a high degree of concavity can cause regions to abruptly enter and exit shadow during local daytime. These instantaneous discontinuities in solar forcing can induce unphysical numerical behaviour within the semi-implicit solver. When applying TEMPEST to highly irregular or concave bodies, the user should therefore be aware that extreme step changes in insolation may reduce numerical stability, and that the explicit solver may be more suitable. One adjustment that can be made to the model is to apply a smoothing operation to the diurnal insolation for each facet. This spreads the solar forcing discontinuity across multiple time steps, leading to more stable behaviour. This should be done with care, however, to ensure the modelled insolation is still physically justified. \textcolor{revtwo}{This is a known feature of the Crank-Nicolson method, which remains stable but responds to a sudden step change in forcing with oscillations that die away over the following time steps (see Subsection~\ref{subsec:Initialisation}).}


\subsection{Comparison with other one-dimensional thermal models}

\begin{figure}
    \centering
    \includegraphics[trim={0 0 0 0},clip, width=\linewidth]{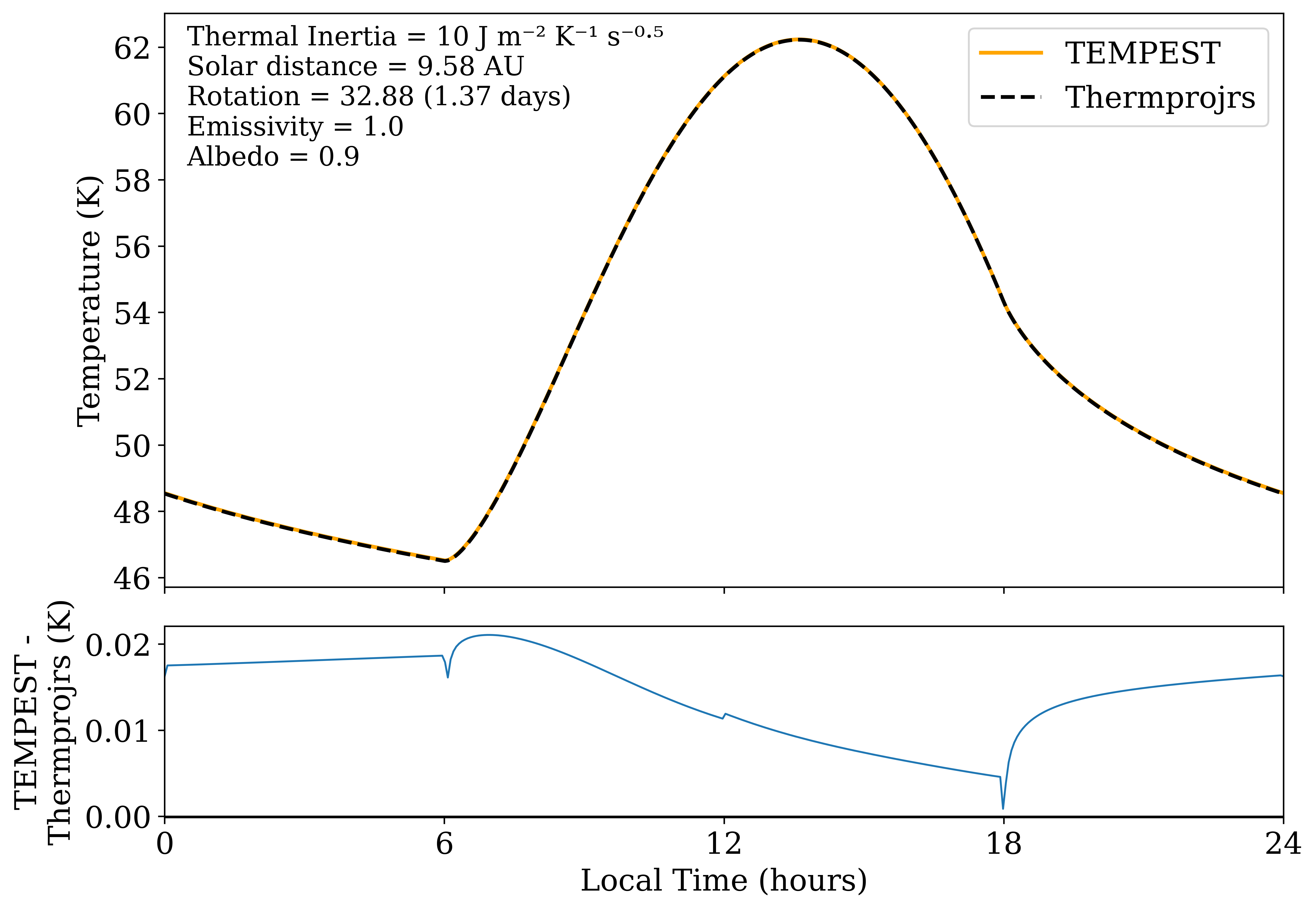}
    \caption{Comparison of diurnal surface temperature predictions from TEMPEST and Thermprojrs for an Enceladus-like setup. The upper panel shows the diurnal temperature as a function of local time over one full rotation, and the lower panel shows the residual (TEMPEST - Thermprojrs). The two models agree to within 0.02\,K across the full diurnal cycle.}
    \label{fig:thermprojrs_comparison}
\end{figure}

The validation runs given in this section were chosen to challenge each of the key stages in producing the map of resultant surface temperatures. Results from the model's one-dimensional solver were compared with Spencer's Thermprojrs one-dimensional model for a set of identical Enceladus-like input parameters, so that differences attributable to the solvers themselves could be isolated. As shown in Figure~\ref{fig:thermprojrs_comparison}, the temperature-time profiles agree to within $\Delta T < 0.02$\,K across the full diurnal cycle, a difference that can be considered negligible for most observing and remote-sensing applications. The slight residual is likely due primarily to the different treatment of the deepest-layer temperature $T_{\rm{deep}}$ at the end of each day: where Thermprojrs applies a nearest-neighbour correction factor to $T_{\rm{deep}}$, TEMPEST sets it to the radiative mean of the surface temperature (Subsection~\ref{subsec:Initialisation}).

\subsection{Comparison with other three-dimensional thermophysical models}

\begin{figure*}
    \centering
    \includegraphics[trim={0 0 0 0},clip, width=\linewidth]{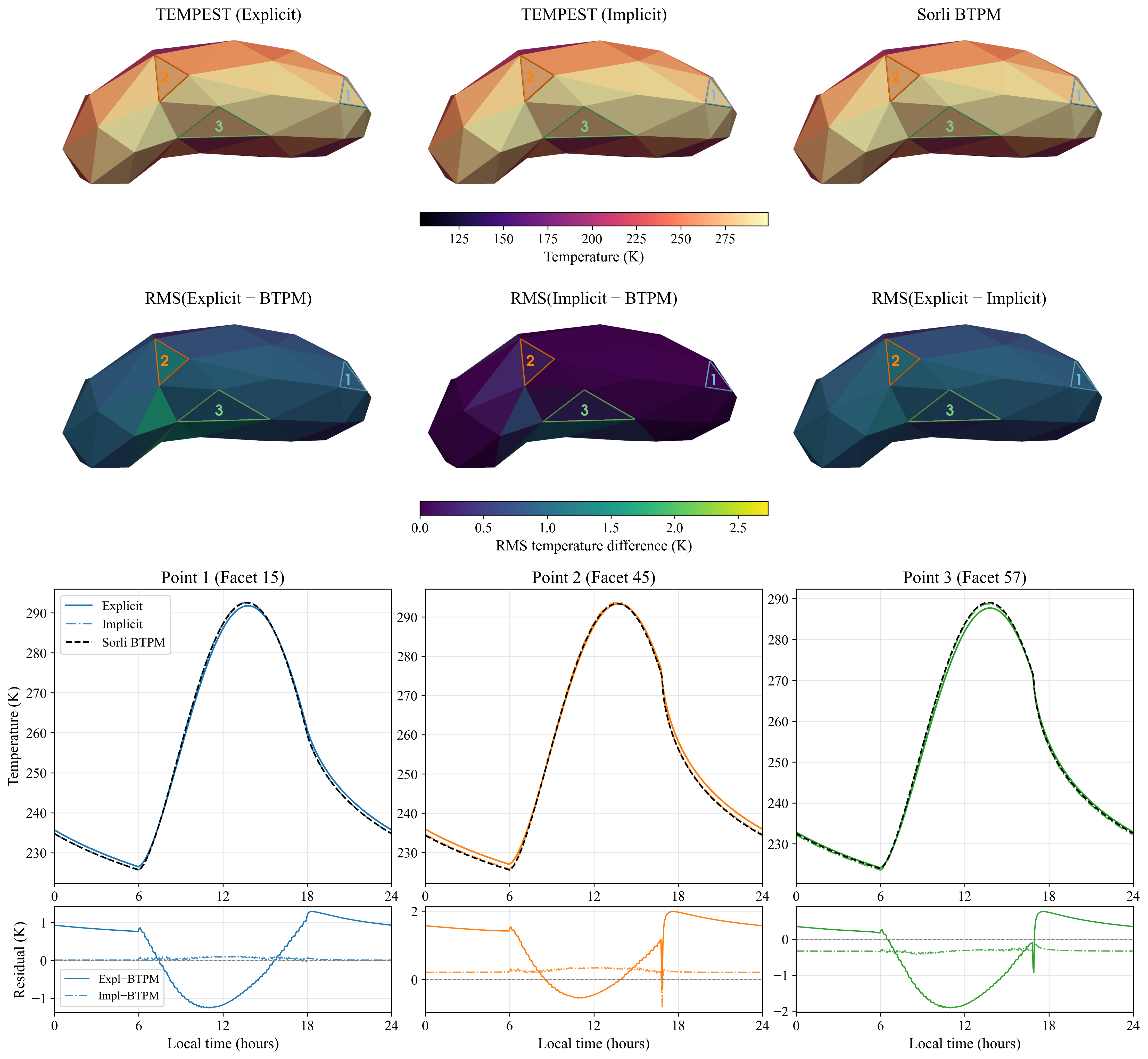}
    \caption{Comparison between TEMPEST using explicit and semi-implicit solvers and the Sorli Binary Thermophysical Model (BTPM) for Itokawa at the same rotational state. \textit{Top two rows:} Surface temperature maps for the explicit (Thermprojrs-like) and semi-implicit (Crank--Nicolson) TEMPEST solvers and the Sorli BTPM (upper row), together with root mean square temperature-difference maps over all time steps. \textit{Bottom two rows:} Diurnal surface temperature curves (upper) and residuals relative to the BTPM (lower) at the three labelled facets, plotted against local solar time with noon at 12\,h. Solid and dash-dot lines show the explicit and semi-implicit TEMPEST solvers, respectively; dashed black lines show the Sorli BTPM. Parameters used in both models were matched to: $\rho = 1190$\,kg\,m$^{-3}$, $c = 750$\,J\,kg$^{-1}$\,K$^{-1}$, $\Gamma = 700$\,J\,m$^{-2}$\,K$^{-1}$\,s$^{-1/2}$ (equivalent to $k \approx 0.549$\,W\,m$^{-1}$\,K$^{-1}$), $\varepsilon = 0.9$, $A = 0.02$, heliocentric distance $r_h = 1.32$\,AU, and rotation period $P = 12.132$\,h. Shadowing and radiative self-heating were enabled; TEMPEST used two scattering iterations. \textcolor{author}{Both models were run for 80 rotations, and the BTPM was run with uniform layer spacing (10 layers per diurnal skin depth, to 20 skin depths).}}
    \label{fig:sorli_comparison}
\end{figure*}

Results from the three-dimensional model were compared with those of the Binary TPM of \citet{sorli2025thermophysical} running as a single body. This comparison was intended to evaluate the treatment of morphology-driven surface effects, namely shadowing, scattering, and radiative self-heating. The Binary TPM was chosen for this in part due its similar modelling approach, which allows a direct comparison, but also because it implements the one-dimensional thermal diffusion equation using a fundamentally different numerical approach. This ensures that the comparison tests the underlying physical formulation rather than simply reproducing results with an equivalent solver. Figure~\ref{fig:sorli_comparison} gives a comparison of the two models. \textcolor{revone}{Both models were run for 80 rotations, and the BTPM was run with uniform layer spacing (10 layers per diurnal skin depth, to 20 skin depths) for consistency with TEMPEST. The semi-implicit solver agrees with the BTPM with a mean difference of 0.07\,K and an RMS difference of 0.46\,K. On facets that receive no radiative self-heating the RMS difference is 0.13\,K, consistent with the $\sim$0.1\,K discretisation error of each model relative to a high-resolution one-dimensional reference solution. The explicit solver, run at its stability-limited default time step, damps the diurnal amplitude by $\sim$0.7\,K (RMS 0.96\,K), which falls to $\sim$0.1\,K with four times as many time steps. The largest differences occur within a few minutes of shadow ingress and egress (up to 6\,K), where the BTPM samples shadowing from a precomputed table at 1$^\circ$ intervals of rotation, and on concave facets (up to $\sim$2.5\,K), where the two models compute view factors differently.}

Figure~\ref{fig:sorli_comparison} shows shadow passing over facets 45 and 57 at almost identical times, with close agreement in $\Delta T$ across the body. This is taken to mean that the geometric treatment of shadowing, light scattering, and radiative self-heating are in agreement between these models. 

Notably, in order to perform the model-to-model comparison in three-dimensions, the BTPM had to be edited to switch off its treatment of temperature-dependent thermal conductivity and specific heat capacity. This was necessary as the one-dimensional solvers would otherwise produce substantially different temperatures when given identical inputs, but means the BTPM is not being used as it is intended, and this should be taken into consideration when looking at the surface temperature plots in Figure~\ref{fig:sorli_comparison}. 


\subsection{Validation against Lunar Reconnaissance Orbiter data in one dimension}

In order to validate the model’s ability to predict surface temperatures under a range of mission-relevant circumstances it was important to find sites that are well observed in the thermal infrared. The Diviner instrument on Lunar Reconnaissance Orbiter (LRO) \citep{paige2010lunar} provides a suitable database for model validation, with four channels in the thermal infrared covering a range of local times across the entire lunar surface. Additionally, the digital elevation models (DEMs) available from LRO's Lunar Orbiter Laser Altimeter (LOLA) provide global coverage of topographic elevation at 128 pixels per degree (ppd) \citep{zuber2010lunar}, providing a suitable shape-model input for TEMPEST.

\begin{figure}
    \centering
    \includegraphics[trim={0 0 0 0},clip, width=\linewidth]{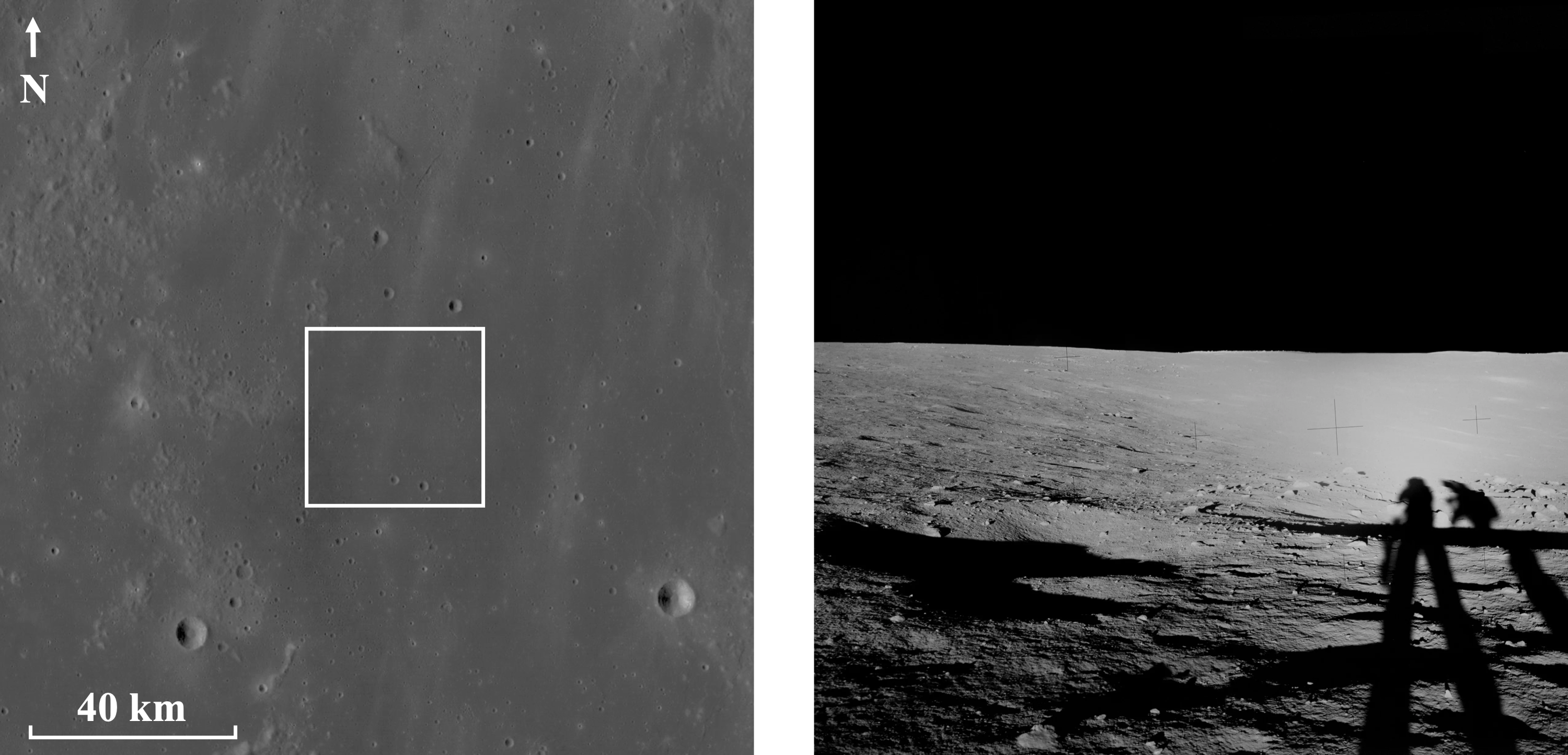}
    \caption{Left: LROC WAC image with the chosen site used for one-dimensional validation highlighted \protect\citep{robinson2010lroc_edr}; the site lies in basaltic mare terrain within Oceanus Procellarum. Right: Apollo 12 panorama of Middle Crescent Crater (within the region of interest), taken by Pete Conrad on 19 November 1969; panorama assembled by Dave Byrne from NASA image sequence AS12-68-6845 to AS12-68-6852.}
    \label{fig:Lunar_mare}
\end{figure}

For a one-dimensional diurnal test designed to focus primarily on the accuracy of the one-dimensional solver over a full diurnal cycle, a region was sought where surface temperatures are not dominated by the effects of scattered light or radiative mutual heating from overlooking terrain (Figure~\ref{fig:Lunar_mare}). This leaves a surface boundary condition that can be approximated as a function of only the solar zenith angle, Bond albedo and local surface slope angle (see e.g. Section 4.2.1 of \cite{williams2017global}  for a detailed discussion). 

\begin{figure}
    \centering
    \includegraphics[trim={0 0 0 0},clip, width=\linewidth]{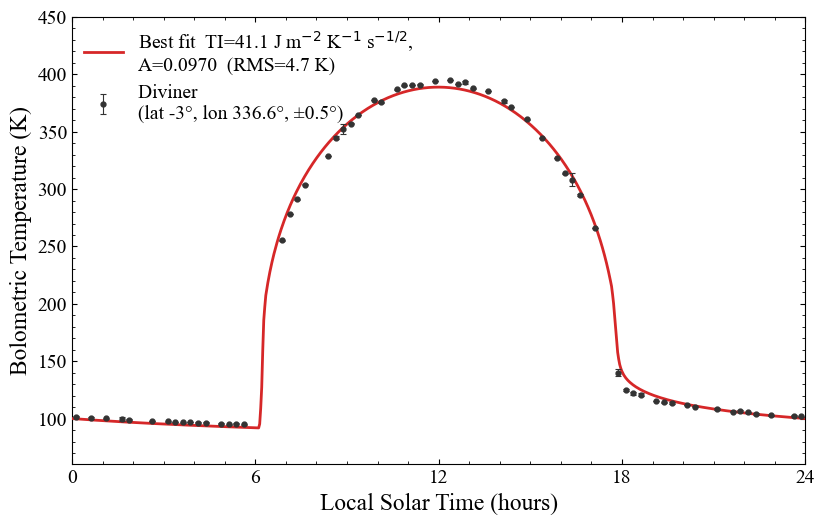}
    \caption{Diurnal curve for the modelled site centred at $3.0^\circ$S, $336.6^\circ$E using Diviner-derived surface temperatures. Black points show Diviner-derived bolometric temperatures binned by local solar time with their associated error, while the red curve shows the best-fitting one-dimensional TEMPEST model. The fitted thermal inertia and Bond albedo are consistent with values expected from regional albedo and H-parameter maps \protect\cite{hayne2017global}.}
    \label{fig:1D_Diviner_validation_plot}
\end{figure}

The Diviner bolometric temperatures for the chosen area were extracted and plotted in 0.25 hour local-time bins to produce the diurnal curve shown in Figure~\ref{fig:1D_Diviner_validation_plot} \textcolor{revone}{(local time divides the lunar day into 24 local hours, each corresponding to $\sim$29.5 Earth hours)}. The radiative mean temperature, $\left(\langle T^4 \rangle\right)^{1/4}$, of the observations in each bin was used, and the standard deviation of measurements within each bin was plotted as the uncertainty. In several bins the derived temperatures are highly consistent, so the error bars of each plot are not always visible.

TEMPEST was run to convergence in its one-dimensional configuration for this site, using the local surface slope and orientation derived from the LOLA DEM and a single representative facet to isolate the one-dimensional solver. The thermal inertia $\Gamma$ and Bond albedo $a$ were both treated as free parameters and varied to minimise the root-mean-square difference between the modelled and Diviner-derived temperatures across the diurnal cycle. The best-fitting solution is shown as the red curve in Figure~\ref{fig:1D_Diviner_validation_plot}, with $\Gamma = $ 41.1 \tiu\ and $a = $ 0.097. These values are consistent with those expected from regional thermal inertia and albedo maps \citep{hayne2017global}, indicating that the one-dimensional solver reproduces the observed diurnal behaviour using physically reasonable surface properties.


\subsection{Validation against LRO data in three-dimensions}

\begin{figure}
    \centering
    \includegraphics[trim={0 0 0 0},clip, width=\linewidth]{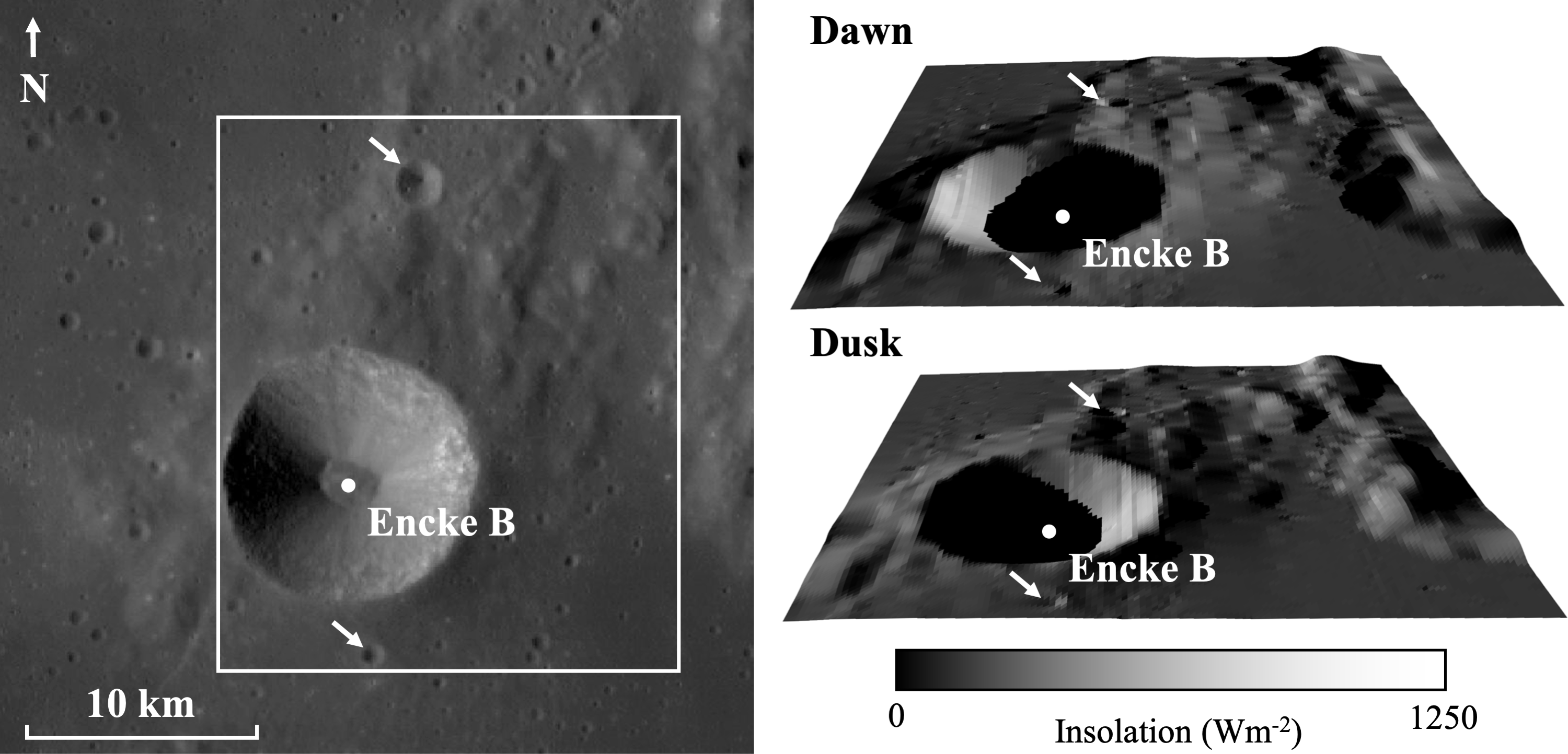}
    \caption{Left: LROC WAC image of the region of terrain chosen for three-dimensional validation with the region of interest highlighted. Right: LRO LOLA digital elevation model of the region of interest shown at dawn and dusk \protect\citep{smith2023lola_edr}. \textcolor{revone}{The perspective views are shaded with dawn and dusk insolation to highlight the relief} by plotting insolation values calculated by TEMPEST as a colour map on terrain facets. The 11\,km crater, Encke B is labelled, with the two smaller craters to the north and south identified by arrows.}
    \label{fig:Lunar_ROI}
\end{figure}

For validation in three dimensions, the second region chosen is a cratered area of lunar mare at $2.55^\circ$ N, $323.4^\circ$ E (Figure~\ref{fig:Lunar_ROI}). This 20\,km square region is large enough that the 128\,ppd LOLA DEM resolution is sufficient to enable terrain effects such as scattering and radiative self-heating on the scale of the Diviner thermal observations to be modelled. The region-of-interest (ROI) contains the 11\,km wide Encke B crater, two smaller craters to its north and south, and an area of elevated terrain in the north-east that are all large enough to be resolved in the DEM. These varied terrains allow for thorough testing of the handling of scattering and radiative self-heating in TEMPEST. Another reason for choosing this ROI was that it does not have substantial high terrain neighbouring the modelled region that would have a radiative influence on the ROI.

\begin{figure*}
    \centering
    \includegraphics[trim={0 0 0 0},clip, width=\linewidth]{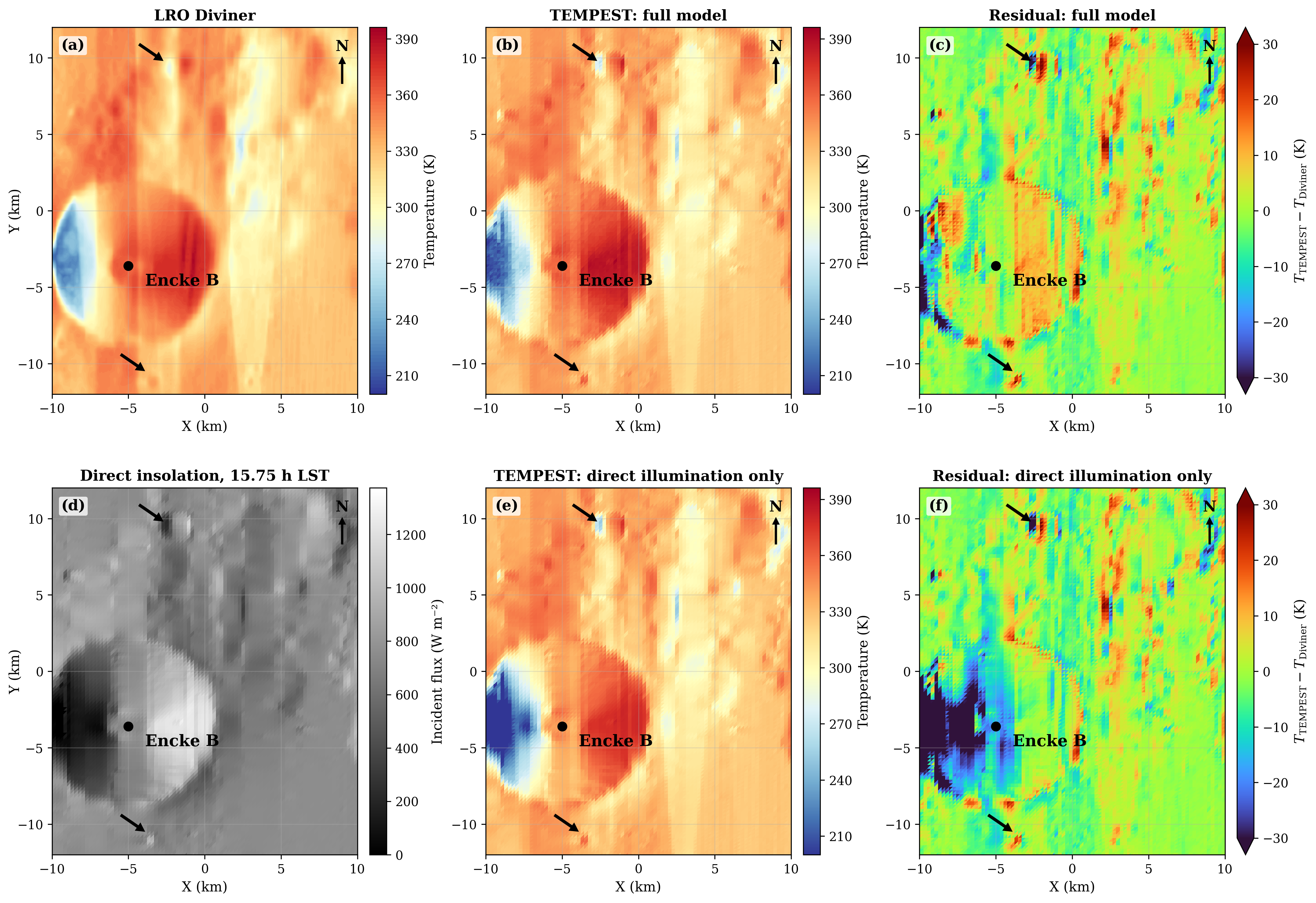}
    \caption{\textcolor{revone}{Comparison of LRO Diviner temperatures with TEMPEST predictions for the site at $2.55^\circ$N, $323.4^\circ$E, in basaltic mare within Oceanus Procellarum, for a window of one local hour (1/24 of a lunar day, $\sim$29.5 Earth hours) centred on a local time of 15.75\,h. (a) Diviner-derived surface temperature. (b) TEMPEST including shadowing, scattering and radiative self-heating, sampled at the local time of the Diviner observations at each point. (c) Residual, (b) minus (a). (d) Direct insolation calculated by TEMPEST at 15.75\,h in the same plan view, with the Sun to the west (left), $\sim$34$^\circ$ above the horizon. (e) TEMPEST with direct illumination and shadowing only (scattering and self-heating disabled). (f) Residual, (e) minus (a). The pointed ends of the residual colour bars indicate that the scale saturates: residuals beyond $\pm$30\,K (0.6\,\% of facets in c, 3.4\,\% in f) are shown in the end colours. Arrows mark the two smaller craters to the north and south of Encke B. The low residuals between the observed and predicted temperature (right) show that the model accurately reproduces temperatures for the flat regions of the terrain selection, and captures a notable level of detail in surface temperature heterogeneity.} 
    }
    \label{fig:TEMPEST_Diviner_maps}
\end{figure*}

To produce the Diviner-derived temperatures map (Figure~\ref{fig:TEMPEST_Diviner_maps}, left), the raw LRO Diviner data covering the ROI were first downloaded from the RDR Query website. The local time window of observations 15.25\,h to 16.25\,h was selected as it contained a large number (48,390) of observations within the ROI across the instrument's four broad thermal channels (6-9, with wavelength ranges $\sim$13–23, $\sim$25–41, $\sim$50–100, and $\sim$100–400\,µm). The data were filtered at the point of download to include only observations with optimal interpolated calibration (QCA = 0), geometry derived from definitive spacecraft pointing and ephemeris tracking data (QGE = 12, bits 2–3), and no instrument anomalies such as eclipse, thermal instability, or actuator motion during acquisition (QMI = 0); all 111,045 records in the downloaded dataset satisfied these criteria.

A 100\,m Cartesian grid representing the ROI was created for each channel. Each grid-point was then populated by taking the Gaussian-weighted mean brightness temperature of nearby Diviner observations within the filtered dataset using a Gaussian kernel matching the nominal 160\,m by 320\,m Diviner spatial footprint, truncated at 3$\sigma$. 

The effective wavelength of each channel was calculated by taking the geometric mean of the upper and lower wavelength bounds. We then found the best fitting brightness temperature across all four thermal maps by comparing the four gridded brightness temperatures with those calculated from a Planck function, assuming emissivity of 0.95 \citep{bandfield2015lunar}. This fitted effective kinetic surface temperature can be directly compared to the TEMPEST thermal map, and is plotted on the left-hand panel of Figure~\ref{fig:TEMPEST_Diviner_maps}. 

\begin{figure}
    \centering
    \includegraphics[trim={0 0 0 0},clip, width=\linewidth]{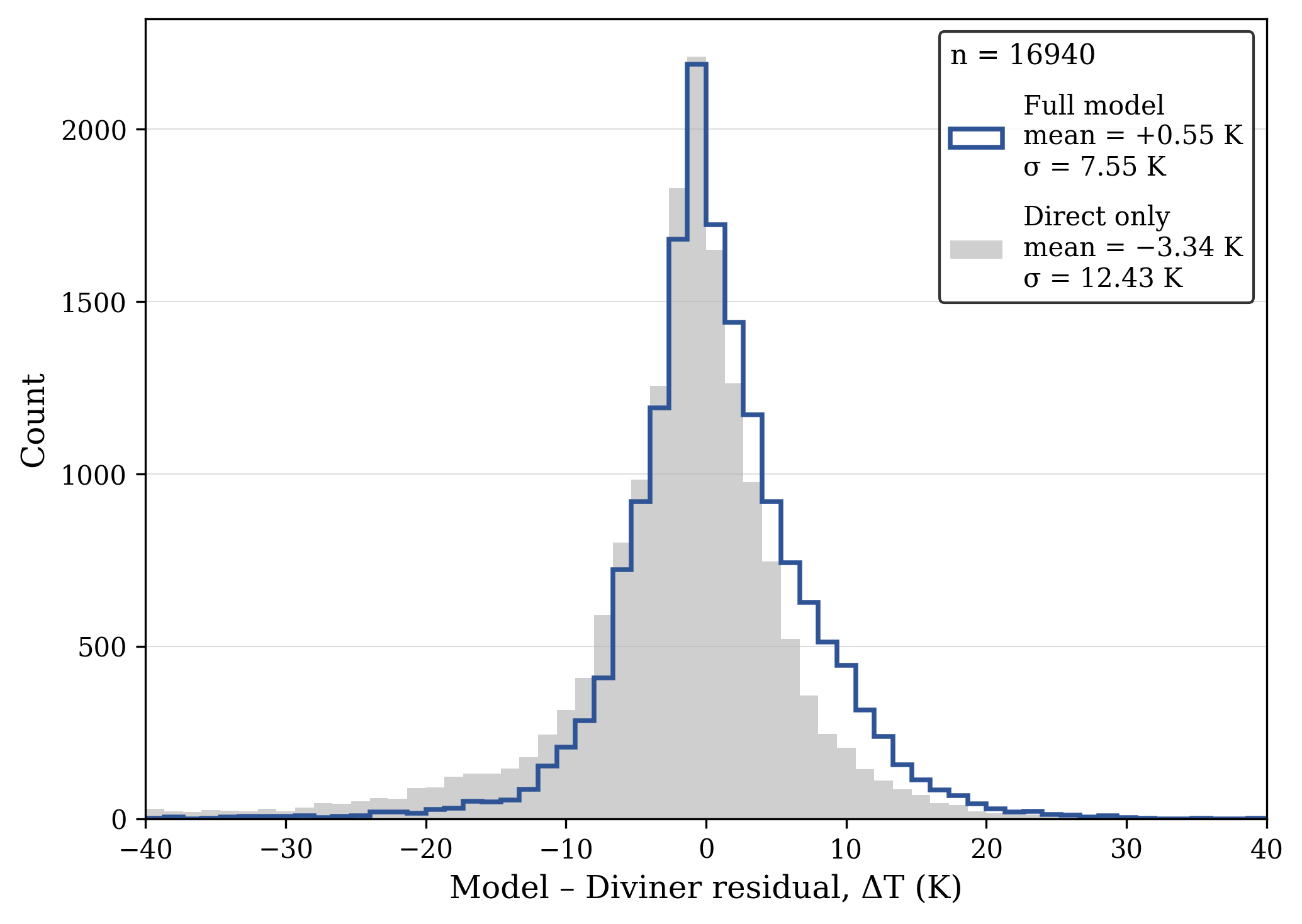}
    \caption{\textcolor{revone}{Distribution of the temperature residuals in Figure~\ref{fig:TEMPEST_Diviner_maps} for the full model (blue outline; panel~c) and with direct illumination only (grey; panel~f).} 
    }
    \label{fig:Validation_histogram}
\end{figure}

To produce the TEMPEST temperature prediction map (Figure~\ref{fig:TEMPEST_Diviner_maps}, centre) the first step was to download the LOLA DEM for the terrain section around the point of interest. This was run in TEMPEST assuming a thermal inertia of 55\tiu, taken from the thermal inertia map produced using thermophysical properties at 273\,K by \cite{hayne2017global}. The Bond albedo used, 0.119, was calculated by taking the mean LOLA 1064\,nm albedo for the ROI \citep{lucey2014global}, and multiplying by the empirically derived thermal scaling factor of $f = 0.49$ from \cite{hayne2017global}. The plotted data are the kinetic temperatures of each facet projected onto a 2-dimensional plane and aligned with the LRO Diviner temperatures map. The panel on the right-hand side shows the difference between the model temperatures and Diviner temperatures ($T_{\rm{TEMPEST}} - T_{\rm{Diviner}}$) on this same grid. Figure~\ref{fig:Validation_histogram} shows the residuals plotted as a histogram; this shows agreement in the mean temperature of the two maps, with the distribution of temperature residuals having \textcolor{revone}{a mean residual of 0.55 K and a standard deviation of 7.55 K.}

The final step was to account for variability in the local times of Diviner observations -- a grid of mean local solar time (LST) of Diviner observations (Gaussian-weighted by the same spatial kernel) was saved alongside the temperature grid. For each TEMPEST facet, after spatial alignment, the Diviner LST at that location was taken, and TEMPEST was sampled at the corresponding local time (with linear interpolation between adjacent time steps) to account for the fact that different parts of the ROI were observed at slightly different times of day. The banding visible in the left and centre panels of Figure~\ref{fig:TEMPEST_Diviner_maps} is an observational artefact of different swaths being observed at different local times. By accounting for variations in local time in this way, TEMPEST reproduces the temperature variation associated with the banding.

\textcolor{revone}{To quantify the contribution of scattering and radiative self-heating, the simulation was repeated with both disabled, so that each facet receives only direct sunlight subject to shadowing (Figure~\ref{fig:TEMPEST_Diviner_maps}, lower row). At the modelled local time the Sun is $\sim$34$^\circ$  above the horizon and only 0.3~\% of facets in the ROI are shadowed (Figure~\ref{fig:TEMPEST_Diviner_maps}d), but indirect heating is nonetheless significant within Encke B. Given the low lunar albedo, it is dominated by thermal emission from the sunlit eastern wall of the crater. It raises temperatures on the east-facing western wall by $\sim$30 K (median) and on the crater floor by $\sim$14 K, while having a negligible effect on the surrounding mare (median 0.3 K). Without these terms the western wall is too cold by a median of 31 K (Figure~\ref{fig:TEMPEST_Diviner_maps}f), compared with 1.5 K when they are included, and the standard deviation of the residuals across the ROI increases from 7.55 K to 12.43 K (Figure~\ref{fig:Validation_histogram}). TEMPEST's treatment of scattering and self-heating is therefore required to reproduce the observed temperatures within the crater.}

This Diviner comparison was performed to test the ability of TEMPEST to reproduce empirical spacecraft data in three-dimensions, and in that respect, TEMPEST performs well. \textcolor{revone}{The two thermal maps show excellent qualitative agreement (<5~K) across 64.1~\% of the ROI,} with some major deviations that are closely linked to geographic features. The key differences between the model and observations are as follows:
\begin{enumerate}
    \item \textcolor{revtwo}{The eastern half of Encke B is substantially warmer in the TEMPEST result than in the Diviner map. The TEMPEST map shows smooth-surface kinetic temperatures, without the roughness correction of Subsection~\ref{sec:spherical_depressions}, whereas the Diviner temperatures are derived from the observed radiance and so include the effect of unresolved roughness. The west-facing eastern wall receives the most direct insolation in the ROI, and so has the largest sub-facet temperature contrasts. With the Sun to the west and Diviner observing close to nadir, the hottest, sun-facing sub-facet slopes are partly hidden from view, so the observed temperature is lower than the smooth-surface kinetic temperature. The surrounding mare is illuminated more obliquely, so the effect is smaller there. A test of the roughness scheme against these data is beyond the scope of this comparison, which was designed to isolate the treatment of shadowing, scattering and self-heating.} Close inspection of the LROC WAC imagery in Figure~\ref{fig:Lunar_ROI} also indicates that there may be patches of high albedo regolith here that are not captured in TEMPEST.
    \item Encke B is cooler on the western rim in the TEMPEST model. This appears to be \textcolor{revone}{due to} the shape-model not fully capturing the sharp edge of the crater rim, meaning the clear boundary seen in the Diviner data is smoothed out. \textcolor{revtwo}{The cooler regions are localised rather than varying smoothly with illumination or viewing geometry, consistent with small-scale artefacts in the shape-model.}    
    \item There is a substantial mismatch in temperatures in the crater at $Y = 10$~km. Close inspection of the shape-model reveals that while this crater is large enough to be captured in the LOLA DEM, its geometry is substantially altered by discretisation of elevation points, and the local slopes and shadowing are not accurately represented. 
    \item North/south striations in Encke B crater appear to be an artefact of the LOLA DEM with some north/south rows of facets having systematically shifted normals from their neighbours. Figure~\ref{fig:Lunar_ROI} shows the crater to have smooth geometry without the linear features seen in the shape-model.
\end{enumerate}

There are a number of additional limitations of this TEMPEST model that we are aware of, though they may not be explicitly visible within Figure~\ref{fig:TEMPEST_Diviner_maps}. Seasonality is not considered in this model run. The Diviner observations used here span 12 passes between 2010 and 2025, covering essentially the full ±1.5$^\circ$ range of lunar sub-solar latitude variation. While this variation can have a substantial effect on polar temperatures, at 2.55$^\circ$\,N the change in solar zenith angle at 15.75\,h local time across the full seasonal cycle produces a surface temperature variation of only $\sim$0.4\,K on flat terrain (computed using the albedo and emissivity values adopted here). \textcolor{revone}{This is well below the 7.55\,K standard deviation of the residuals,} so the decision not to include a seasonal correction is justified for this equatorial site.

Another limitation of TEMPEST not highlighted in this section is that it is not currently possible to model regions with spatially varying Bond albedo or thermal inertia, so in the test cases the model had to be run once for each of these. \textcolor{revtwo}{This is not a fundamental limitation, as the energy balance is solved independently for each facet, and support for spatially varying properties is under development.} The sites used for validation were chosen in part due to their limited regional variation of such properties, so the impact of this on scattered light and thermal inertia is minimal. It would, however, present issues on targets with more significant albedo variegation e.g., Europa, so applications to these \textcolor{revtwo}{bodies should consider this by only modelling small regions,} or larger areas where variegation in physical properties is minimal. 

A final important consideration is that temperature dependence of thermal inertia has been understood to be an important consideration on planetary regoliths, particularly those experiencing major diurnal temperature variation, as noted since the Apollo era \citep[e.g.][]{LINSKY1966606}. The two solvers built into TEMPEST (explicit and semi-implicit) do not account for temperature-varying properties, and as such, some variation between Diviner temperatures, and those predicted by TEMPEST was expected. A valuable extension of this work would be integration of a thermal solver with temperature and depth dependent thermophysical properties, for example, the heat1d model of \cite{hayne2017global}. 


\subsection{Performance benchmarking}
\label{subsec:benchmarking}

In order to benchmark the model, it was run for two different shape-models at a range of resolutions on two different platforms: 

\begin{enumerate}
    \item \textbf{Personal laptop:} a 2022 Apple MacBook Pro (M2, 8-core CPU, 8\,GB RAM, macOS 15)
    \item \textbf{Server:} an institutional server (2$\times$ Intel Xeon Gold 6240R, 48 cores / 96 threads, 376\,GB RAM, Debian 12)
\end{enumerate}

TEMPEST was configured with \texttt{n\_jobs = 8} and \texttt{32}, respectively. The Gaskell Itokawa shape-model \citep{gaskell2021itokawa} was obtained from NASA's Planetary Data System (PDS) at 3,145,728 facets and downsampled using the decimate tool in Blender. The Bennu shape-model \citep{daly2020hemispherical} was obtained from the NASA Scientific Visualization Studio\footnote{\url{https://svs.gsfc.nasa.gov/5069}} and downsampled in the same way.

\begin{figure}
    \centering
    \includegraphics[trim={0 0 0 0},clip, width=0.8\linewidth]{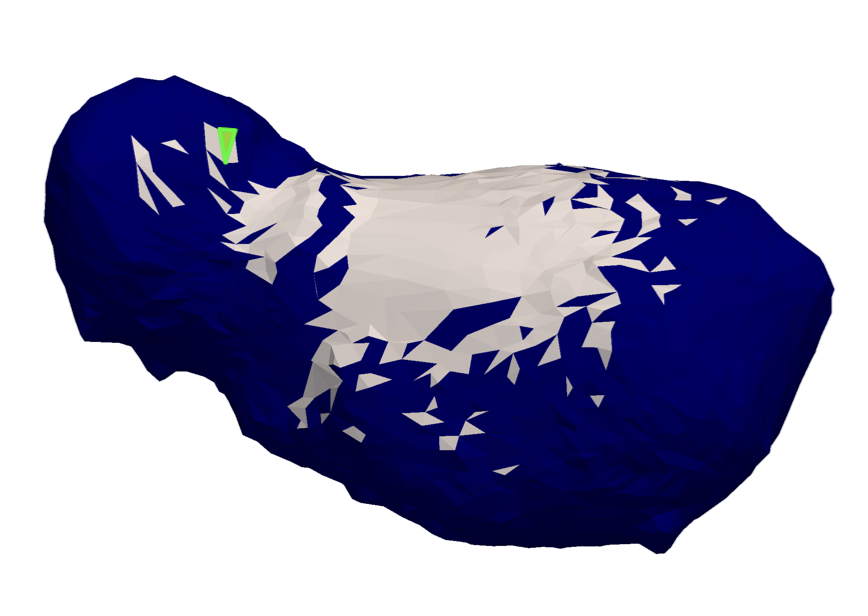}
    \caption{The bilobate nature of 25143 Itokawa presents a challenge for many models due to the large number of facets visible to each other. A test facet is shown above in green, with all of the facets visible to it shown in white.}
    \label{fig:concavity}
\end{figure}

Bennu and Itokawa were chosen to illustrate the impact of concavity on computational performance. Bennu, being a primarily convex shape, runs substantially faster than Itokawa at the same resolution (number of facets). Concave geometries such as Itokawa do increase runtime, but the Monte Carlo ray-tracing approach keeps this overhead manageable by sampling a representative subset of the visible opposing terrain rather than evaluating every possible facet interaction (Figure~\ref{fig:concavity}). This allows the thermal and scattering contributions of the opposing lobe to be captured without the much larger computational cost of a fully exhaustive treatment. The mean number of facets visible to each other facet is 13.57 for the 3,184-facet Bennu model and 57.96 for the 3,144-facet Itokawa model. The greatest difference in computation time was observed in the scattering and thermal solver processes, which scale with the inter-facet visibility requirements.

\begin{figure}
    \centering
    \includegraphics[trim={0 0 0 0},clip, width=\linewidth]{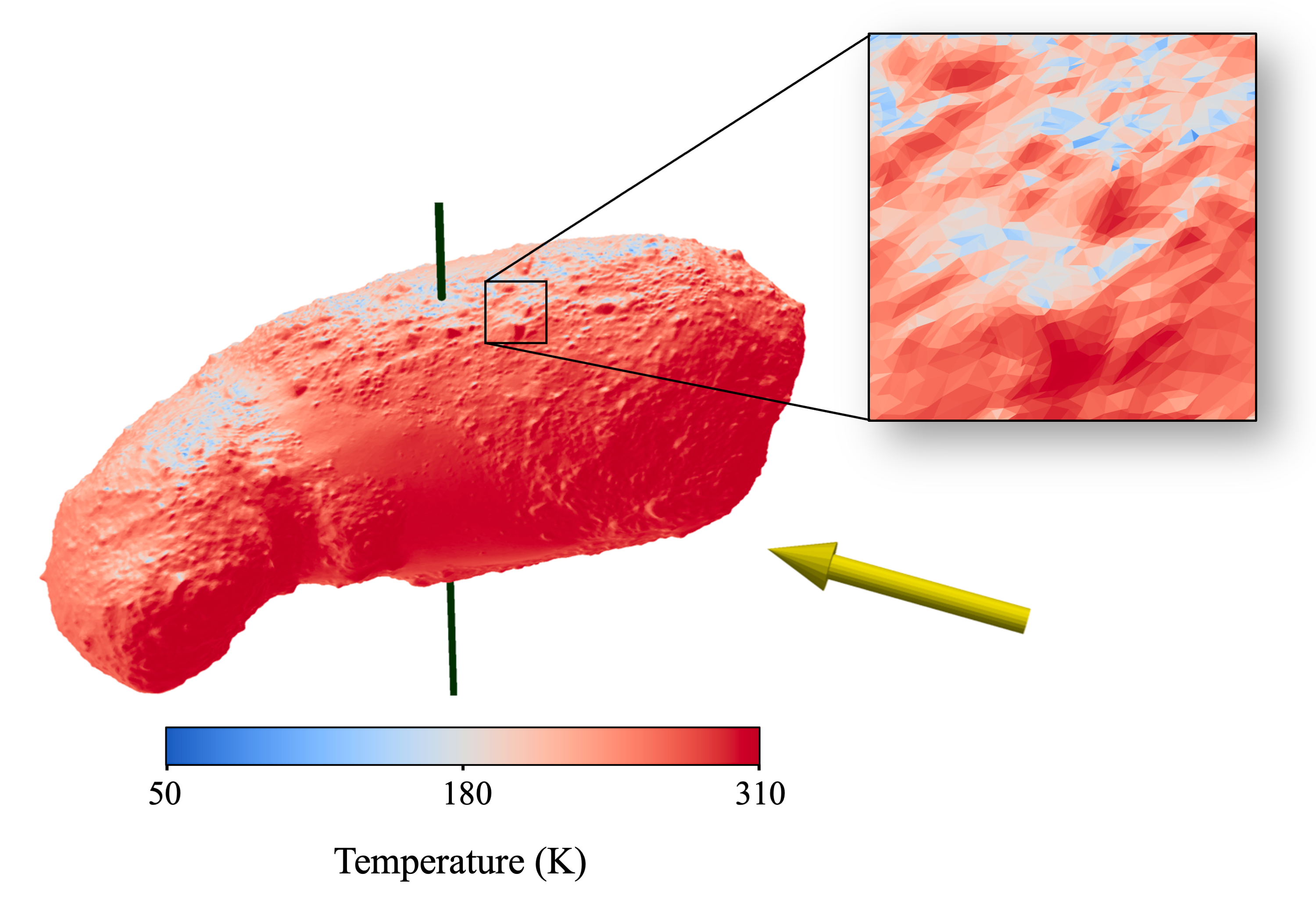}
    \caption{The high resolution 314,571 facet model of Itokawa used for benchmarking. An enlarged section shows the scale of individual facets. The sunlight direction is given in yellow and the axis of rotation is shown in green. This model was run on the server described in Subsection~\ref{subsec:benchmarking} using the semi-implicit solver. }
    \label{fig:itokawa_high_res}
\end{figure}

\begin{table*}
\caption{Performance benchmarking for TEMPEST. Wall-clock times are given in HH:MM:SS.S (0.1\,s precision). Convergence criterion was $\Delta \rm{T_{mean}}$ < 0.1\,K in all cases. All runs used Itokawa thermophysical parameters. View factors are cached and therefore computed only during the first run for a given shape-model.}

\label{tab:benchmark}
\centering
\scriptsize
\setlength{\tabcolsep}{4pt}
\renewcommand{\arraystretch}{1.15}
\textbf{Explicit solver}

\begin{tabular}{l c cccc @{\hspace{2.5em}} cccc}
\hline
 &  & \multicolumn{4}{c}{Laptop} & \multicolumn{4}{c}{Server} \\
\cline{3-10}
shape-model & Facets 
 & View Factors & Scattering & Thermal Solver & Total
 & View Factors & Scattering & Thermal Solver & Total \\
\hline
Itokawa (super low-res) & 314 
 & 00:00:13.8 & 00:00:10.4 & 00:00:03.6 & 00:00:27.8
 & 00:00:39.4 & 00:00:22.5 & 00:00:05.4 & 00:01:07.3 \\

Itokawa (low-res) & 3,144 
 & 00:00:38.7 & 00:00:19.1 & 00:00:58.0 & 00:01:55.8
 & 00:00:39.0 & 00:00:26.1 & 00:00:32.9 & 00:01:38.0 \\

Itokawa (mid-res) & 31,456 
 & 00:21:08.0 & 00:33:00.0 & 00:10:11.0 & 01:04:19.0
 & 00:09:57.0 & 00:11:16.0 & 00:09:21.0 & 00:30:34.0 \\

Bennu (low-res) & 3,184 
 & 00:00:40.6 & 00:00:06.7 & 00:00:21.9 & 00:01:09.2
 & 00:00:39.8 & 00:00:07.4 & 00:00:32.2 & 00:01:19.4 \\
\hline
\end{tabular}

\vspace{1em}

\textbf{Semi-implicit solver}

\begin{tabular}{l c cccc @{\hspace{2.5em}} cccc}
\hline
 &  & \multicolumn{4}{c}{Laptop} & \multicolumn{4}{c}{Server} \\
\cline{3-10}
shape-model & Facets 
 & View Factors & Scattering & Thermal Solver & Total
 & View Factors & Scattering & Thermal Solver & Total \\
\hline
Itokawa (super low-res) & 314 
 & 00:00:13.8 & 00:00:07.8 & 00:00:01.3 & 00:00:22.9
 & 00:00:39.4 & 00:00:22.4 & 00:00:04.2 & 00:01:06.0 \\

Itokawa (low-res) & 3,144 
 & 00:00:38.7 & 00:00:14.9 & 00:00:45.2 & 00:01:38.8
 & 00:00:39.0 & 00:00:18.5 & 00:00:18.2 & 00:01:15.7 \\

Itokawa (mid-res) & 31,456 
 & 00:21:08.0 & 00:24:00.0 & 00:07:21.0 & 00:52:29.0
 & 00:09:57.0 & 00:11:35.0 & 00:08:51.0 & 00:30:23.0 \\

Itokawa (high-res) & 314,571 
 & \multicolumn{4}{c}{---- Insufficient RAM ----}
 & 02:24:54.6 & 07:51:36.9 & 25:55:04.2 & 36:11:35.7 \\

Bennu (low-res) & 3,184 
 & 00:00:40.6 & 00:00:04.9 & 00:00:17.6 & 00:01:03.1
 & 00:00:39.8 & 00:00:05.9 & 00:00:16.9 & 00:01:02.6 \\
\hline
\end{tabular}

\end{table*}

As the table shows, a TEMPEST run is split into 3 key stages, each taking a similar amount of time to complete. In the case of a parameter space search where a model is being run repeatedly with varied thermal inertia, for example, the results of the view factor and scattering calculation steps can be reused, so only the thermal solver is run each time. The explicit and semi-implicit solvers take a similar amount of time to run per time step, but for primarily convex shapes the semi-implicit solver is more stable with a smaller number of time steps. For this reason for the benchmarking runs in Table~\ref{tab:benchmark} the explicit solver was run with 360 time steps per rotation, while the semi-implicit solver was run with just 240, leading to improved performance in most cases. The main disadvantage of the semi-implicit solver is that it does not handle discontinuities in illumination conditions well, and these can lead to unstable behaviour. For this reason, the user is advised to select the explicit solver when modelling bodies with a high degree of concavity, e.g. bilobate asteroids, or deep lunar terrains. TEMPEST is capable of running very large shape-models, with 314,571 facet shape-model of Itokawa shown in Figure~\ref{fig:itokawa_high_res} given \textcolor{revtwo}{as an example. However, these models} are computationally expensive, so for most purposes reducing the spatial resolution of the model or modelling only a small section of the whole body is preferred. A key point to note is that in many cases the laptop runs are actually faster than the server. This is mainly the case for small models (<10,000 facets), however, where parallel overheads dominate runtimes. 


\section{Summary and Conclusions}
\label{Conclusions}

The newly released TEMPEST thermophysical model puts the latest developments in thermophysical modelling techniques into a single open-source model designed to balance ease of use with availability of features. This model is the first open-source TPM known to the authors to combine non-Lambertian scattering with a configurable unresolved-roughness treatment. The model has been validated in the one-dimensional case against Spencer's Thermprojrs \cite{SPENCER1989337}, in three dimensions against Sorli's BTPM \cite{sorli2025thermophysical}, and directly against mission data of the Moon from the Lunar Reconnaissance Orbiter in one and three dimensions. These validation tests show good agreement with published models and the ability to reproduce remote-sensing observations from spacecraft. 

As with any thermophysical model, accurate results depend on configuring the solver and shape-model appropriately for the problem at hand, and here we highlight some key considerations. The semi-implicit solver is stable at larger time steps and is the faster choice for smooth or convex bodies; however, for highly concave geometries such as bilobate asteroids or craters, where sharp shadow transitions can drive unphysical oscillations, the explicit solver is the appropriate choice and handles these cases robustly. At low resolutions, the centroid-based view-factor and shadowing calculations do not consider partial shadowing of individual facets. This shortcoming can be addressed by subdividing facets or using a shape-model with a higher resolution. Users should also be aware of the limits of the input shape-model itself; gridding of elevation data can introduce facet-scale artefacts, as seen in the LOLA-derived DEM of Section~\ref{sec:Validation}. Finally, TEMPEST currently assumes spatially uniform thermophysical properties, neglects seasonal variation in sub-solar latitude, and does not capture the temperature dependence of thermal inertia. These can introduce systematic errors on bodies where the corresponding effect is significant. The modularity of the one-dimensional solver means that in future the model can be adapted to new targets by pairing it with one-dimensional solvers that can handle the different physical effects that dominate in different regimes, e.g. non-uniform surface layering, temperature dependent thermophysical properties, and the solid-state greenhouse effect. A core motivation for development of TEMPEST was to contribute to the study of the icy moons of the outer Solar System. Understanding these thermal environments requires models that incorporate endogenic heating, eclipses, seasonal variability, and spatially varying albedo. Extending TEMPEST in these directions comes with the opportunity to couple surface thermal models with deeper subsurface models, linking surface temperature observations with ocean and ice shell dynamics to better constrain the energy budget and potential habitability.


\section*{Acknowledgements}

We thank J. R. Spencer, J. P. Emery, P. O. Hayne, J. Penn and K. Sorli for their insights and helpful discussions throughout the development of TEMPEST and the preparation of this manuscript. 

DGL thanks the Science and Technology Facilities Council (STFC) for a PhD studentship (ST/Y509474/1).

\section*{Data Availability}
 
The full code, documentation, video demonstrations, and example model products are freely available to download at github.com/duncanLyster/TEMPEST under an MIT Attribution License. The code can be adapted to suit specific needs, and the authors are happy to collaborate and advise on projects using TEMPEST. All shape-models used in this paper can be downloaded from the PDS. 

\section*{Conflict of Interest Statement }

\textcolor{editor}{The authors declare no conflict of interest.}



\bibliographystyle{mnras}
\bibliography{bibliography} 

@article{2006Spencer,
       author = {Spencer, J. R.},
       title = "{Cassini Encounters Enceladus: Background and the Discovery of a South Polar Hot Spot}",
       journal = {Science},
       volume = {311},
       number = {5766},
       pages = {1401-1405},
       year = 2006,
       month = mar,
       doi = {10.1126/science.1121661}
}

@article{howett2011high,
  title={High heat flow from Enceladus' south polar region measured using 10--600 cm- 1 Cassini/CIRS data},
  author={Howett, CJA and Spencer, JR and Pearl, J and Segura, M},
  journal={Journal of Geophysical Research: Planets},
  volume={116},
  number={E3},
  year={2011},
  publisher={Wiley Online Library}
}

@article{SPENCER1989337,
title = {Systematic biases in radiometric diameter determinations},
journal = {Icarus},
volume = {78},
number = {2},
pages = {337-354},
year = {1989},
issn = {0019-1035},
doi = {https://doi.org/10.1016/0019-1035(89)90182-6},
url = {https://www.sciencedirect.com/science/article/pii/0019103589901826},
author = {John R Spencer and Larry A Lebofsky and Mark V Sykes}
}

@article{spencer1990rough,
  title={A rough-surface thermophysical model for airless planets},
  author={Spencer, John R},
  journal={Icarus},
  volume={83},
  number={1},
  pages={27--38},
  year={1990},
  publisher={Elsevier}
}

@article{harris2002asteroids,
  title={Asteroids in the thermal infrared},
  author={Harris, Alan W and Lagerros, Johan SV},
  journal={Asteroids III},
  volume={205},
  year={2002},
  publisher={Univ. of Arizona Press Tucson}
}

@phdthesis{curtis2022measuring,
  title={Measuring and utilising visible light scattering functions for the lunar regolith using the visible Oxford space environment goniometer},
  author={Curtis, Rowan},
  year={2022},
  school={University of Oxford}
}

@article{paige2010lunar,
  title={The lunar reconnaissance orbiter diviner lunar radiometer experiment},
  author={Paige, DA and Foote, MC and Greenhagen, BT and Schofield, JT and Calcutt, S and Vasavada, AR and Preston, DJ and Taylor, FW and Allen, CC and Snook, KJ and others},
  journal={Space Science Reviews},
  volume={150},
  number={1},
  pages={125--160},
  year={2010},
  publisher={Springer}
}

@article{williams2017global,
  title={The global surface temperatures of the Moon as measured by the Diviner Lunar Radiometer Experiment},
  author={Williams, J-P and Paige, DA and Greenhagen, BT and Sefton-Nash, E},
  journal={Icarus},
  volume={283},
  pages={300--325},
  year={2017},
  publisher={Elsevier}
}

@article{zuber2010lunar,
  title={The lunar reconnaissance orbiter laser ranging investigation},
  author={Zuber, Maria T and Smith, David E and Zellar, Ronald S and Neumann, Gregory A and Sun, Xiaoli and Katz, Richard B and Kleyner, Igor and Matuszeski, Adam and McGarry, Jan F and Ott, Melanie N and others},
  journal={Space Science Reviews},
  volume={150},
  number={1},
  pages={63--80},
  year={2010},
  publisher={Springer}
}

@article{lucey2014global,
  title={The global albedo of the Moon at 1064 nm from LOLA},
  author={Lucey, PG and Neumann, GA and Riner, MA and Mazarico, E and Smith, DE and Zuber, MT and Paige, DA and Bussey, DB and Cahill, JT and McGovern, A and others},
  journal={Journal of Geophysical Research: Planets},
  volume={119},
  number={7},
  pages={1665--1679},
  year={2014},
  publisher={Wiley Online Library}
}

@article{bandfield2015lunar,
  title={Lunar surface roughness derived from LRO Diviner Radiometer observations},
  author={Bandfield, Joshua L and Hayne, Paul O and Williams, Jean-Pierre and Greenhagen, Benjamin T and Paige, David A},
  journal={Icarus},
  volume={248},
  pages={357--372},
  year={2015},
  publisher={Elsevier}
}

@article{augier2021reducing,
  title={Reducing the ecological impact of computing through education and Python compilers},
  author={Augier, Pierre and Bolz-Tereick, Carl Friedrich and Guelton, Serge and Mohanan, Ashwin Vishnu},
  journal={Nature Astronomy},
  volume={5},
  number={4},
  pages={334--335},
  year={2021},
  publisher={Nature Publishing Group UK London}
}

@article{delbo2015asteroid,
  title={Asteroid thermophysical modeling},
  author={Delbo, Marco and Mueller, Michael and Emery, Joshua P and Rozitis, Ben and Capria, Maria Teresa},
  journal={Asteroids iv},
  volume={1},
  pages={107--128},
  year={2015},
  publisher={University of Arizona Press Tucson}
}

@inproceedings{moller2005fast,
author = {M\"{o}ller, Tomas and Trumbore, Ben},
title = {Fast, minimum storage ray/triangle intersection},
year = {2005},
isbn = {9781450378338},
publisher = {Association for Computing Machinery},
address = {New York, NY, USA},
url = {https://doi.org/10.1145/1198555.1198746},
doi = {10.1145/1198555.1198746},
booktitle = {ACM SIGGRAPH 2005 Courses},
pages = {7–es},
location = {Los Angeles, California},
series = {SIGGRAPH '05}
}

@techreport{schuckart2025modular,
  title={Modular heat transfer code for dry, airless planetary surfaces: MoCSI},
  author={Schuckart, Christian and Aussel, Ben and R{\"u}ckriemen-Bez, Tina and G{\"u}ttler, Carsten and Gundlach, Bastian},
  year={2025},
  institution={Copernicus Meetings}
}

@article{sorli2025thermophysical,
title = {A 3D thermophysical model for binary asteroid systems: Application to the BYORP effect on (175706) 1996 FG3},
journal = {Icarus},
volume = {434},
pages = {116527},
year = {2025},
issn = {0019-1035},
doi = {https://doi.org/10.1016/j.icarus.2025.116527},
url = {https://www.sciencedirect.com/science/article/pii/S0019103525000740},
author = {Kya C. Sorli and Paul O. Hayne and Rachel H. Cueva and Chloe J. Long and Jay W. McMahon and Daniel J. Scheeres},
}

@inproceedings{henry2024thermal,
  title={Thermal Infrared Imager (TIRI) Simulation with Kalast for Hera Flyby Targets},
  author={Henry, Gr{\'e}goire and Karatekin, {\~A} and Berk Senel, Cem and others},
  booktitle={Europlanet Science Congress 2024},
  pages={EPSC2024--1167},
  year={2024}
}

@article{wohlfarth2023advanced,
  title={An advanced thermal roughness model for airless planetary bodies-Implications for global variations of lunar hydration and mineralogical mapping of Mercury with the MERTIS spectrometer},
  author={Wohlfarth, Kay and W{\"o}hler, Christian and Hiesinger, Harald and Helbert, J{\"o}rn},
  journal={Astronomy \& Astrophysics},
  volume={674},
  pages={A69},
  year={2023},
  publisher={EDP Sciences}
}

@inproceedings{kanamaru2022development,
  title={Development of Thermophysical Model to Simulate Non-Gravitational Acceleration on Binary Asteroid Didymos and Dimorphos},
  author={Kanamaru, Masanori and Tommei, Giacomo and Okada, Tatsuaki and Sakatani, Naoya and Shimaki, Yuri and Tanaka, Satoshi and Ishizaki, Takuya and Senshu, Hiroki and Arai, Takehiko and Sekiguchi, Tomohiko and others},
  booktitle={AGU Fall Meeting Abstracts},
  volume={2022},
  pages={NH11A--01},
  year={2022}
}

@article{king2020oxford,
  title={The Oxford 3D thermophysical model with application to PROSPECT/Luna 27 study landing sites},
  author={King, Oliver and Warren, Tristram and Bowles, Neil and Sefton-Nash, Elliot and Fisackerly, Richard and Trautner, Roland},
  journal={Planetary and Space Science},
  volume={182},
  pages={104790},
  year={2020},
  publisher={Elsevier}
}

@article{hanuvs2015thermophysical,
  title={Thermophysical modeling of asteroids from WISE thermal infrared data--Significance of the shape model and the pole orientation uncertainties},
  author={Hanu{\v{s}}, Josef and Delbo, Marco and {\v{D}}urech, Josef and Al{\'\i}-Lagoa, Victor},
  journal={Icarus},
  volume={256},
  pages={101--116},
  year={2015},
  publisher={Elsevier}
}

@article{kieffer2013thermal,
  title={Thermal model for analysis of Mars infrared mapping},
  author={Kieffer, Hugh H},
  journal={Journal of Geophysical Research: Planets},
  volume={118},
  number={3},
  pages={451--470},
  year={2013},
  publisher={Wiley Online Library}
}

@article{muller2012surface,
  title={Surface properties of asteroids from mid-infrared observations and thermophysical modeling},
  author={M{\"u}ller, Michael},
  journal={arXiv preprint arXiv:1208.3993},
  year={2012}
}

@article{rozitis2011directional,
  title={Directional characteristics of thermal--infrared beaming from atmosphereless planetary surfaces--a new thermophysical model},
  author={Rozitis, Ben and Green, Simon F},
  journal={Monthly Notices of the Royal Astronomical Society},
  volume={415},
  number={3},
  pages={2042--2062},
  year={2011},
  publisher={Blackwell Publishing Ltd Oxford, UK}
}

@article{delbo2004nature,
  title={The Nature of Near Earth Asteroids From the Study of Their Thermal Infrared Emission (Doctoral Dissertation)},
  author={Delb{\'o}, M},
  journal={Freie University},
  volume={557},
  year={2004}
}

@article{lagerros1998thermal,
  title={Thermal physics of asteroids. IV. Thermal infrared beaming},
  author={Lagerros, Johan SV},
  journal={Astronomy and Astrophysics, v. 332, p. 1123-1132 (1998)},
  volume={332},
  pages={1123--1132},
  year={1998}
}

@book{hapke2012theory, 
    place={Cambridge}, 
    edition={2}, 
    title={Theory of Reflectance and Emittance Spectroscopy}, 
    publisher={Cambridge University Press}, 
    author={Hapke, Bruce}, 
    year={2012}
}

@article{hapke1998opposition,
    title = {The Opposition Effect of the Moon: Coherent Backscatter and Shadow Hiding},
    journal = {Icarus},
    volume = {133},
    number = {1},
    pages = {89-97},
    year = {1998},
    issn = {0019-1035},
    doi = {https://doi.org/10.1006/icar.1998.5907},
    url = {https://www.sciencedirect.com/science/article/pii/S0019103598959073},
    author = {Bruce Hapke and Robert Nelson and William Smythe}
}

@article{hapke1963photometric,
  title={Photometric studies of complex surfaces, with applications to the Moon},
  author={Hapke, Bruce and van Hoen, Hugh},
  journal={Journal of Geophysical Research},
  volume={68},
  number={15},
  pages={4545--4570},
  year={1963},
  publisher={Wiley Online Library}
}

@inbook{LeVeque_2002, 
    place={Cambridge}, 
    series={Cambridge Texts in Applied Mathematics}, 
    title={Convergence, Accuracy, and Stability}, 
    booktitle={Finite Volume Methods for Hyperbolic Problems}, 
    publisher={Cambridge University Press}, 
    author={LeVeque, Randall J.}, 
    year={2002}, 
    pages={139–157}, 
    collection={Cambridge Texts in Applied Mathematics}
}

@article{hayne2017global,
  title={Global regolith thermophysical properties of the Moon from the Diviner Lunar Radiometer Experiment},
  author={Hayne, Paul O and Bandfield, Joshua L and Siegler, Matthew A and Vasavada, Ashwin R and Ghent, Rebecca R and Williams, Jean-Pierre and Greenhagen, Benjamin T and Aharonson, Oded and Elder, Catherine M and Lucey, Paul G and others},
  journal={Journal of Geophysical Research: Planets},
  volume={122},
  number={12},
  pages={2371--2400},
  year={2017},
  publisher={Wiley Online Library}
}

@article{braga2026surface,
  title={Surface Properties, Orbital Dynamics, and Thermophysical Modeling of the Primitive Asteroid (269) Justitia},
  author={Braga, Leonardo and Amarante, Andre and Ferreira, Alessandra and Monteiro, Filipe and Martins, Maria},
  journal={The Planetary Science Journal},
  volume={7},
  number={2},
  pages={46},
  year={2026},
  publisher={The American Astronomical Society}
}

@techreport{maclennan2024thermal,
  title={Thermal Emission Modeling of Rough Surfaces for Comparison to Radar Observations},
  author={MacLennan, Eric and Virkki, Anne},
  year={2024},
  institution={Copernicus Meetings}
}

@unpublished{howes2026thermal,
  author = {Howes, Sarah E. and Howett, C. J. A. and Lyster, D. G. and Lange, L. and Piqueux, S.},
  title  = {Thermal Inertia and Bolometric Bond Albedo Measurements of Europa's Surface using Galileo PPR},
  year   = {2026},
  note   = {Submitted to MNRAS},
}

@article{LINSKY1966606,
title = {Models of the lunar surface including temperature-dependent thermal properties},
journal = {Icarus},
volume = {5},
number = {1},
pages = {606-634},
year = {1966},
issn = {0019-1035},
doi = {https://doi.org/10.1016/0019-1035(66)90075-3},
url = {https://www.sciencedirect.com/science/article/pii/0019103566900753},
author = {Jeffrey L. Linsky}
}

@article{sorli2026thermal,
  title={Thermal segregation and reddening in Europa's double ridges},
  author={Sorli, Kya C and Hayne, Paul O and Lange, Lucas and Piqueux, Sylvain},
  journal={Journal of Geophysical Research: Planets},
  volume={131},
  number={5},
  pages={e2025JE009408},
  year={2026},
  publisher={Wiley Online Library}
}

@article{jorda2016global,
  title={The global shape, density and rotation of Comet 67P/Churyumov-Gerasimenko from preperihelion Rosetta/OSIRIS observations},
  author={Jorda, Laurent and Gaskell, R and Capanna, C and Hviid, S and Lamy, Philippe and {\v{D}}urech, J and Faury, G and Groussin, Olivier and Guti{\'e}rrez, P and Jackman, C and others},
  journal={Icarus},
  volume={277},
  pages={257--278},
  year={2016},
  publisher={Elsevier}
}

@article{daly2020hemispherical,
  title={Hemispherical differences in the shape and topography of asteroid (101955) Bennu},
  author={Daly, MG and Barnouin, OS and Seabrook, JA and Roberts, J and Dickinson, C and Walsh, KJ and Jawin, ER and Palmer, EE and Gaskell, R and Weirich, J and others},
  journal={Science Advances},
  volume={6},
  number={41},
  pages={eabd3649},
  year={2020},
  publisher={American Association for the Advancement of Science}
}

@article{davidsson2015interpretation,
  title={Interpretation of thermal emission. I. The effect of roughness for spatially resolved atmosphereless bodies},
  author={Davidsson, Bj{\"o}rn JR and Rickman, Hans and Bandfield, Joshua L and Groussin, Olivier and Guti{\'e}rrez, Pedro J and Wilska, Magdalena and Capria, Maria Teresa and Emery, Joshua P and Helbert, J{\"o}rn and Jorda, Laurent and others},
  journal={Icarus},
  volume={252},
  pages={1--21},
  year={2015},
  publisher={Elsevier}
}

@book{alma990194379310107026, 
    author = {Press, William H. and Teukolsky, Saul A. and Vetterling, William T. and Flannery, Brian P.}, 
    title = {Numerical Recipes in C: The Art of Scientific Computing}, 
    edition = {2nd}, 
    publisher = {Cambridge University Press}, 
    address = {Cambridge}, 
    year = {1992}, 
    isbn = {0521431085}
}

@misc{gaskell2021itokawa,
  author       = {Gaskell, Robert W. and Saito, Jun and Ishiguro, Masateru and Kubota, Takashi and Hashimoto, Takashi and Hirata, Naoyuki and Abe, Seiji and Barnouin, Olivier S. and Scheeres, Daniel J.},
  year         = {2021},
  title        = {Gaskell Itokawa Shape Model V1.1},
  publisher    = {NASA Planetary Data System},
  note         = {urn:nasa:pds:gaskell.ast-itokawa.shape-model::1.1},
  doi          = {10.26033/3b2j-yy57},
  url          = {https://sbn.psi.edu/pds/resource/doi/itokawashape_1.1.html}
}

@misc{robinson2010lroc_edr,
  author       = {Robinson, M. S.},
  title        = {{Lunar Reconnaissance Orbiter Camera Experimental Data Record}},
  year         = {2010},
  howpublished = {{NASA Planetary Data System}},
  note         = {{LRO-L-LROC-2-EDR-V1.0}}
}

@misc{smith2023lola_edr,
  author       = {Smith, D. E. and Neumann, G. A.},
  title        = {{Lunar Reconnaissance Orbiter LOLA Raw Data (EDR) Bundle}},
  year         = {2023},
  publisher    = {{NASA Planetary Data System}},
  doi          = {10.17189/spm2-kv61},
  url          = {https://pds.nasa.gov/ds-view/pds/viewBundle.jsp?identifier=urn:nasa:pds:lro_lola_edr&version=1.0},
  note         = {{urn:nasa:pds:lro_lola_edr::1.0}}
}









\bsp	
\label{lastpage}
\end{document}